\documentstyle[eqsecnum,aps,epsf]{revtex}                        

\draft

\twocolumn

\title{Beyond Uncertainty:\\
the internal structure of electrons and photons}

\author{W. A. Hofer} 
\address{Technische Universit\"at Wien\\
         A--1040 Vienna, Austria}

\begin{document}
\maketitle


\begin{abstract} 
The wave--structure of moving electrons is analyzed on a fundamental level 
by employing a modified de Broglie relation. Formalizing the wave--function 
$\psi$ in real notation yields internal energy components due to mass
oscillations. The wave--features can then be referred to physical waves of 
discrete frequency $\nu$ and the classical dispersion relation 
$\lambda \,\nu = u $, complying with the classical wave equation. 
Including external potentials yields the Schr\"odinger equation, which, in
this context, is arbitrary due to the internal energy components. It can
be established that the uncertainty relations are an expression of this,
fundamental, arbitrariness. Electrons and photons can 
be described by an identical formalism, providing formulations equivalent to
the Maxwell equations. The wave equations of intrinsic particle properties 
are Lorentz invariant considering total energy of particles, although 
transformations into a moving reference frame lead to an increase of 
intrinsic potentials. Interactions of photons and electrons are treated 
extensively, the results achieved are equivalent to the results in quantum 
theory. Electrostatic interactions provide, a posteriori, a justification
for the initial assumption of electron--wave stability: the stability of 
electron waves can be referred to vanishing intrinsic fields of interaction. 
The concept finally allows the conclusion that a significant 
correlation for a pair of spin particles in EPR--like
measurements is likely to violate the uncertainty relations.
\end{abstract} 



\section{Introduction}

I would like to revive, from a new viewpoint, a somewhat old fashioned 
discussion, which recently seems to be all but extinct. As goes
without saying, every contribution, correction, criticism or
suggestion is greatly appreciated.  

The discussion, which lasted for  the major part of the early
stages of atomic, nuclear and particle physics, as we know them
today, dealt primarily with the scientific and logical implications
of quantum theory. Again revived by David Bohm's
contribution, the ``hidden variables'', it gradually began to loose on
impact, until today any fundamental question in this respect seems to
be considered a lack of scientific soundness. 

From an epistemological point
of view, the fact is not easily understood, since all the fundamental
problems are still far from any solution. Second quantization, on the
contrary, provides an even more problematic conception by its
inherent infinity problems \cite{SCH94},
which, till now, have not been solved in a satisfying way. 

That all the original problems are still there, can be seen by a
survey of late publications of Bohm \cite{BOH57}, de Broglie
\cite{BRO57}, Dirac \cite{DIR84}, or Schr\"odinger \cite{SCH92}.

From the viewpoint of common sense the most profound reason of 
uneasiness with quantum theory, despite its indisputable
successes, has been expressed by Richard Feynman:
{\em You see, my physics students don't understand it either.
That is, because I don't understand it. Nobody does.}
\cite{FEY88}

That the discussion, mentioned above, is nevertheless 
non--existent, might have its origins therefore not so much
in a satisfying status quo, but rather in a lack of alternative
concepts. From an epistemological point of view a well
working theory, despite all its conflicts with logical
reasoning, is better than no theory. The discussion then
leads to the question, why so far no promising concept has
been put forth. Comparing with other sciences, the situation
seems very unusual: seventy years of continuous
research -- from Schr\"odinger's first publications 
\cite{SCH26} -- have neither altered the principles of
quantum theory, nor have they removed the axiomatic qualities
of its fundamental statements. This situation basically leaves
two possibilities: either to accept, that we cannot understand,
i.e. that we {\em have to believe}, or to search for
alternatives.

The attempts so far, to remove the axiomatic qualities of
quantum theory, which can be summarized by the {\em Copenhagen
interpretation} \cite{KOP26}, mainly focussed on the 
statistical qualities attributed to fundamental physical
events. It was thought, that a suitable theoretical framework
could prove quantum events to be integral results of basically
continuous processes. 

The first of these attempts can be 
attributed to Erwin Schr\"odinger \cite{SCH26,HEI77}. His wave
mechanics, now a fundamental part of quantum theory, does not
allow for an interpretation of wave--functions as 
{\em physical} waves due to the contradiction a spreading
wave--packet presents to mass conservation. On a fundamental
level, the contradiction originates in the difference between
phase velocity and group velocity of de Broglie waves 
\cite{LDB22}. Since Schr\"odinger's equation is based on
exactly these qualities of de Broglie waves, it cannot be
retained without rejecting the interpretation of wave--functions
as physical waves. The problem was solved by Born's
interpretation \cite{BORN26}
of $|\psi|^2$ as a {\em probability amplitude},
an interpretation which constitutes an integral part of the
Copenhagen interpretation.

This interpretation of the wave--function is not without its
inherent problems, as Einstein, Podolksky and Rosen pointed out
\cite{EPR35}. If quantum theory contained all the physical information
about a spin--system, thus their line of reasoning, 
then information about a measurement process must be exchanged
with a velocity faster than c, if spin conservation is to be
retained. This leaves, as they point out, only two alternatives:
either quantum theory contradicts the theory of relativity, or
it does not contain all the information about the system:
in this case, quantum theory is incomplete. The scientific 
controversy was subject to much discussion, but never
conclusively solved. Recently, its implications have been
extended by the phenomenon of ''interaction--free'' measurements
\cite{ZEI95}, based on a thought experiment by Renninger
\cite{REN60} and realized by a Mach--Zehnder interferometer. 
Interaction free measurements have different 
implications in a classical or quantum context: while a
measurement without interaction in classical physics does not
alter the system, it leads to a reduction of the wave function
in quantum theory. According to the solution, proposed by Heisenberg
(see ref.\cite{REN60}) only the reduction of the
wave--function constitutes physical reality: in this case the
question is legitimate, why a wave function, obviously devoid
of any physical meaning, plays such an important role in current
theories.

David Bohm's famous theory of ''hidden variables'' \cite{BOH52}
was a far more elaborate concept along the same lines. He tried 
to prove, that the formalism of quantum theory allows for a
sub--layer of hidden variables, and that the framework therefore is,
as Einstein suspected, incomplete. The modification was rejected
due to the work of Bell, who devised a simple situation, where this
assumption could be put to a test \cite{BEL66}. The ''Bell
inequalities'' proved that, within the framework of quantum theory,
such an assumption leads to logical contradictions. The theory of
hidden variables was therefore rejected.

The combined evidence of all these attempts allows the
conclusion, that a modification of quantum theory within its 
fundamental axioms, yielding a framework of which
the quantum aspect will be but a result, is not very promising.
At best, it reproduces numerical results by a different 
mathematical structure, at worst, it not even accounts for
basic experiments. There remain, therefore, three fundamental
questions to be answered prior to any attempted modification
of micro physics:

\begin{itemize}
\item
On what theoretical or experimental basis could an alternative
formulation be based?
\item
Which theoretical concept of physics, statistical, field--theory,
or mechanics, shall be employed?
\item
What constitutes a successful alternative to current formulations ?
\end{itemize}

Experimental evidence exists, that moving electrons exhibit
wave--like features. These qualities have been established beyond
doubt by diffraction experiments of Davisson and Germer 
\cite{DAV22}. Quantum theory cannot directly access wave like
qualities, because, as already mentioned, the periodic
wave function needs to be subjected to a complicated evaluation
process in order to extract its physical relevance. If the initial
problems of wave dispersion and velocity can be overcome, the
experimental result of wave characteristics of moving mass and
its mathematical formalization may well constitute the foundation
of a different theoretical conception. The only other experimental
evidence at hand with equivalent significance, the quantum phenomena,
do not as easily fit into a physical interpretation, since the
basic obstacle, why discrete values should be the only ones 
measured, has to be removed: quantum phenomena are therefore
more remote in terms of basic concepts.

Concerning the theoretical context, it has to be considered, that
quantum theory originally started as a modification of mechanics.
Most of the concepts of the theory are consequently still
disguised mechanical concepts. Field theoretical results are
therefore only achieved by a statistical superstructure: the
mechanical basis then leads to a statistical formulation of
field theories, which allows for particle interactions, but not
for internal particle structures. Modifications of this
methodology basically occur along two lines: either by a 
modification of interactions, which requires additional
qualities of particles, or by postulating the existence of
sub--particles. Physical progress then requires an ever refined 
toolkit of particles, sub--particles and interactions. From the
viewpoint of epistemology, the problem inherent to this approach
is the limited degree of freedom. Every modification requires
a major change of particle properties, and since every
new experiment provides data not yet accounted for, the 
development of physics becomes an ever increasing inventory of
individual particles. The ''particle--zoo'', high energy physicists
so frequently refer to, can, along these epistemological lines,
be seen as a consequence of theoretical shortcomings.

Since the experimental basis -- the wave characteristics and
thus the internal structure of particles -- suggest an approach
based on field theory rather than mechanics, this essay will be
based on field theory. If mechanical 
concepts are required, they always refer to continuous distributions
of physical qualities, which yield discrete mass or charge 
values only by integration.

If one is to try an alternative route of theoretical development,
it seems important, not to get carried away and extend a concept
beyond its limits of proper application. To this end, a goal has
to be defined as well as careful checks of experimental evidence.
A theory, which does not allow for falsification, remains useless,
the same applies to a new theory, which does not deviate from
the current one by at least one significant result. That a new
theory cannot -- and should not -- reproduce all results of an
existing one, goes without saying. Personally, I think that the
following results are sufficient to justify serious consideration 
of a new approach: 

\begin{itemize}
\item
The reproduction of fundamental results of early quantum
theory
\item
The formalization of internal particle structures and the 
proof of physical significance of these structures
\item
A model of (hydrogen) atoms consistent with the formalism
developed and consistent with experimental evidence
\item
The basics of a theory  of particle--interaction  due to
intrinsic characteristics
\item
Provision of at least one significant deviation which
allows for an experimentum crucis
\end{itemize}

The fundamental relations and concepts of this new approach have
already been published or are in print in the {\it Speculations
in Science and Technology} (Chapman and Hall, London)
\cite{HOF95,HOF96}. The accounts
given in this and follow--up papers are somewhat enhanced and 
thorougher versions, making use of the greater freedom of length. 
As already emphasized, any feedback is greatly appreciated and 
readers are welcomed to contact me directly via email at the
email address: whofer@eapa04.iap.tuwien.ac.at


\section{Wave structure}

To allow for internal characteristics of moving particles
we base the following section on the experimental evidence of
wave--features. Theoretically, the framework will be
based on a suitable adaptation of the fundamental relations
found by L. de Broglie. 
Originally, de Broglie's formalization of waves \cite{LDB22}
was based on the framework of Special Relativity, the amplitude of 
a ''de Broglie wave'' given by:

\begin{equation}\label{ws001}
        \psi (x_{\mu}) = \psi_{0} 
        \exp\,- i \left(k^{\mu} x_{\mu}\right)
\end{equation}

The variables are four--dimensional vectors, their components 
are described by the following relations:

\begin{equation}\label{ws002}
        x_{\mu} = (ct, - x, - y, - z) \qquad 
        k^{\mu} = \frac{m \gamma}{\hbar} 
        \left(c, u_{x}, u_{y}, u_{z}\right) 
\end{equation}

In a three dimensional and non--relativistic ($\gamma$ = 1)
adaptation of the original equation (\ref{ws001}) the amplitude
can be expressed as a plane wave. This adaptation is quite common
in quantum theory and therefore no deviation from 
the original framework:

\begin{equation}\label{ws003}
        \psi (\vec x, t) = \psi_{0} \exp \, i 
        \left(\frac{m \, \vec u}{\hbar} \vec x - \omega t\right)
\end{equation}
        
The first major deviation from the traditional framework is the 
interpretation of $ \psi (\vec x, t) $ as a real function, as a
physical wave. This conception is not compatible with quantum
theory for the following reasons: as a moving particle cannot
consist of a specific plane wave (particle mass in this case
would be distributed over an infinite region of space), a moving
particle in quantum theory {\em must} be formalized as a Fourier
integral over partial waves. An application of Schr\"odinger's
equation then leads to the, above mentioned, spreading of the
wave--packet, which contradicts experimental evidence. The 
interpretation of the wave--function as a {\it physical entity},
furthermore, contradicts Heisenberg's uncertainty relation 
\cite{HEI26}, because it implies the possibility of measurements
below quantum level.

\begin{equation}\label{ws004}
        \psi (\vec x,t) = \psi_{0} \sin
        \left(\frac{m \, \vec u}{\hbar} \vec x - \omega t\right)
        \qquad \psi_{0} \in \, R
\end{equation}

The reason, that this--real--formalization of de Broglie remained
so far unconsidered, is not only its inconsistency with quantum
theory, but also a consequence of Planck's relation: dispersion
of de Broglie waves is described by two independent relations, which
are widely confirmed in experiments. The first is the wavelength
of de Broglie waves, established by diffraction experiments of
Davisson and Germer \cite{DAV22}, the second Planck's energy 
relation \cite{PLA01}:

\begin{equation}\label{ws005}
        \lambda (\vec p) = \frac{h}{|\vec p|} \qquad
        E (\omega) = \hbar\,\omega
\end{equation}

If particle energy is classical {\em kinetic} or relativistic energy,
the phase velocity of a de Broglie wave is not equivalent to
particle velocity:

\begin{equation}\label{ws006}
        \hbar \omega = E \quad \Longrightarrow \quad
        c_{phase} = \frac{\omega}{k} = 
        \left\{\frac{c^2}{u}\right\}_{rel} =
        \left\{\frac{u}{2}\right\}_{non-rel}
\end{equation}

The deduction of phase velocity, leading to the contradiction
in Eq. (\ref{ws006}) is, contrary to a first impression, not
self--evident. It contains, on thorougher analysis, a conjecture
and a subsequent logical circle. Firstly, it is not self--evident,
that on a micro--level energy of a particle is restricted to
{\em kinetic} energy of inertial mass: assuming, that a particle is, 
basically, a mechanical entity without any internal energy component, 
cannot be inferred from experimental or theoretical results. It is 
therefore an {\em assumption} lacking a solid basis. Based on 
this conjecture is the following logical circle:

{\em If energy is kinetic energy, phase velocity of a de Broglie
wave is not equal to mechanical velocity of a particle. If phase
velocity does not equal mechanical velocity, a free particle
cannot consist of a single wave of specific frequency. If it
does not consist of a single wave, then the wave--features of a
particle must be formalized as a Fourier integral over infinitely
many partial waves. In this case any partial wave cannot be 
interpreted as a physical wave. If the interpretation as a 
physical wave is not justified, then internal wave--features 
cannot be related to physical qualities. If they cannot be related 
to physical qualities, then internal processes must remain 
unconsidered. And if internal processes remain unconsidered,
then energy of the particle is kinetic energy}

Apart from the question, whether this assumption is necessary, 
which can only be answered a posteriori, it seems also legitimate
to ask, whether it is the best possible assumption, especially in
view of complications arising from fundamental particles 
formalized as Fourier integrals over infinitely many partial
waves. From the viewpoint of simplicity, the concept is
necessary, if no alternative exists, it is unnecessary complicated,
if the same theoretical result -- the wave features of single
particles -- can be obtained by a simpler formalism. To this
aim alternatives have to be analyzed. A direct way of formalization
would be based on  two basic statements: (1) the experimental 
evidence of wave features, and (2) the assumption, that wave
features must be reflected by {\em physical} qualities of particles,
density for example. Mathematically, these statements are expressed
by:

\begin{equation}\label{ws007}
        \psi (\vec x, t) = 
        \psi_{0} \sin \left(\frac{m \vec u}{\hbar}
        \vec x - \omega t\right) 
\end{equation}
\begin{equation}\label{ws008}
        \rho (\vec x, t) = C \, \psi_{0}^2 \sin^2
        \left(\frac{m \vec u}{\hbar}
        \vec x - \omega t\right)
\end{equation}

In view of the dispersion relations Eq. \ref{ws005} these relations
seem contradictory: the standard evaluation procedure would, for a
specific {\em physical} wave, lead to a difference between phase
velocity and mechanical velocity in conflict with the energy 
principle.

\subsection{Kinetic and potential energy}

This contradiction depends, though, on the principal conjecture,
that a moving particle is equivalent to inertial mass, and that
all its energy is contained in the -- longitudinal -- motion of
an inertial mass $m$ with velocity $ |\vec u |$. As soon as wave 
features are assumed to describe the physical nature of particle 
motion, this conjecture can no longer be sustained: energy of the
particle will then be contained in its {\em mass oscillations}. 

Since mass oscillations determine a periodic change of kinetic
energy, the energy principle requires the existence of an 
{\em intrinsic} potential of particle motion, and averaging over
kinetic and potential energy during particle motion then leads to
a total energy of double the original kinetic value. Using this
total energy for Planck's relation satisfies the dispersion 
relation of a monochromatic plane wave: based on the intrinsic
properties of physical waves any monochromatic plane particle
wave then satisfies the wave equation. Averaging density of the
particle wave over a full period we get initially:

\begin{eqnarray}\label{ws009}
        \int_{0}^{\lambda} \rho(\vec x,t = 0) 
        d|\vec x| = \frac{C}{2} \, \psi_{0}^2 \lambda =: \bar{\rho}\,\lambda 
        \nonumber \\
        \bar{\rho} =  \frac{C}{2} \, \psi_{0}^2
\end{eqnarray}

And the kinetic energy of particle motion is therefore
($V_{P}$ denotes the volume of the particle):

\begin{eqnarray} \label{ws010}
        W_{K} = V_{P} \, \frac{\bar{\rho}}{2} |\vec u|^2 =
        \frac{m}{2}\,|\vec u|^2
\end{eqnarray}

Since kinetic energy is not constant but periodic, 
the result requires that an intrinsic potential exists, 
which yields an equivalent average of potential energy:

\begin{eqnarray} \label{ws011}
        W_{P} = V_{P} \, \frac{\bar{\rho}}{2} |\vec u|^2 =
        \frac{m}{2}\,|\vec u|^2
\end{eqnarray}

And total energy of the particle wave is then:

\begin{eqnarray} \label{ws012}
        W_{T} = W_{K} + W_{P} = m\,|\vec u|^2 =: \hbar\,\omega
\end{eqnarray}

Phase velocity of this particle wave equals, consistent with
quantum theory, mechanical velocity of particle motion, although,
contrary to quantum theory, no Fourier integral over partial waves 
is required. Therefore any monochromatic
plane particle wave, consisting of oscillating mass and a complementary
potential, describes motion of a {\em free particle}:

\begin{eqnarray} \label{ws013}
        c_{ph} = \frac{\omega}{|\vec k|} =  \lambda \, \nu =
        \frac{m\,|\vec u|^2}{m\,|\vec u|} = |\vec u|
\end{eqnarray}

From a theoretical point of view the existence of intrinsic 
potentials is a {\em conjecture}, it is more problematic than
the original one based on inertial mass, since it implies the
existence of physical variables not accounted for in the current
framework of quantum theory. Relating $ \hbar \omega $ to total
energy of the particle is, in this respect, not problematic,
since experimentally energy of a particle can only be
estimated by its state of motion or wavelength: and the relation 
between particle velocity $\vec u$, wavelength $\lambda$,
and frequency $\omega$ is retained. 

\subsection{Wave equation}

With the dispersion relation given by Eq. \ref{ws013} the wave function
$\psi(\vec x,t)$ and density $\rho(\vec x,t)$ of free particles 
comply with a wave equation:

\begin{eqnarray} \label{ws014}
        \triangle\, \psi (\vec x,t) - \frac{1}{c_{ph}^2}
        \frac{\partial^2 \psi(\vec x,t)}{\partial t^2} = 0 \\
        \triangle\, \rho (\vec x,t) - \frac{1}{c_{ph}^2}
        \frac{\partial^2 \rho(\vec x,t)}{\partial t^2} = 0   
\end{eqnarray}

The wave equations display intrinsic features of
particle motion and possess, as will be seen in the following sections,
the same dimensional level as electrodynamic formulations.
The logical structure of physical statements is therefore modified:
while in the standard procedure of quantum field theory the 
fundamental statements of quantum theory determine events on a
micro level, and electromagnetic fields are logical superstructures,
in the present context statements about the wave features of particles
are different aspects of the same physical phenomena dealt within
electrodynamics. 

This parallel development of electrodynamics and
quantum theory from the same root -- the intrinsic wave features of
particles -- must be understood in a literal sense, because the
potentials deduced from mass oscillations will be related to
electromagnetic fields.

\subsection{Schr\"odinger equation}

The Schr\"odinger equation, in this context, is an expression
of mechanical features -- i.e. potential and kinetic energy --
of particle motion. From the viewpoint of physical waves it is
not without a certain degree of arbitrariness. To derive the
relation we proceed from the periodic wave function Eq. \ref{ws013}.
Then the time differential is given by:

\begin{eqnarray}\label{ws015}
        |\vec u|^2 \,\triangle \psi = \frac{\partial^2 \psi}{\partial t^2}=
        - \omega^2\, \psi
\end{eqnarray}

Kinetic energy $W_{K}$ of particle motion will therefore be:

\begin{eqnarray}\label{ws016}
        W_{K} \psi = \frac{m}{2}\,|\vec u|^2\,\psi = 
        \frac{\hbar}{2} \,\omega\,\psi
\end{eqnarray}

And substituting Eq. \ref{ws015} into \ref{ws016} 
we get for the kinetic energy of the particle:

\begin{eqnarray} \label{ws017}
        \left( W_{K} + \frac{\hbar^2}{2 m} \triangle \right) \psi = 0 
        \nonumber \\
        W_{K} = - \frac{\hbar^2}{2 m} \triangle 
\end{eqnarray}

If the total energy of a particle is equal to kinetic energy and
a potential $V(\vec r)$:

\begin{eqnarray}\label{ws018}
        W_{T} = V(\vec r) + W_{K}
\end{eqnarray}

then the wave equation in the presence of an external potential
is described by:

\begin{eqnarray}\label{ws019}
        \left(- \frac{\hbar^2}{2 m} \triangle + V(\vec r) \right) \psi = 
        W_{T}\, \psi
\end{eqnarray}

The relation is equivalent to a time--free Schr\"odinger equation
\cite{SCH26}. It is important to consider, that  this (mechanical)
formulation of the wave equation explicitly relates physical potentials
to kinetic energy and thus the wave--length of a material wave (as a
result of the Laplace operator acting on $\psi$) and that, therefore,
in a region, where $V(\vec r)$ takes successively different values,
every kinetic energy and thus wave--length is generally possible.

Its interpretation in the context of material waves is not trivial, though.
If the volume of a particle is finite, then wavelength and frequency
become {\em intrinsic} variables of motion, which implies, due to 
energy conservation, that the wave function itself is a measure
for the potential at an arbitrary point $\vec r$. If it is a measure
for the physical conditions of the environment -- i.e. the intensity
of $V(\vec r)$ --, then it {\em must} have physical relevance.  But if
it has physical relevance, then the EPR dilemma \cite{EPR35} will be
fully confirmed, and the result will favor Einsteins's interpretation
of quantum theory: quantum theory cannot be complete under the condition
that (1) particles do have finite dimensions, and (2) the 
Schr\"odinger equation is a correct description of the variations of
particle waves in the presence of external fields. The only alternative,
which leaves quantum theory intact, is the assumption of particles
with zero volume: an assumption which leads, in the context of 
electrodynamics, to the equally awkward result of infinite energy of
particles.
The first major result of this essay can therefore be formulated 
as follows: 
\begin{itemize}
\item
The interpretation of particles as physical waves
allows the conclusion, that quantum theory cannot be, in a logical
sense, a complete theoretical description of micro physical systems.
\end{itemize}

There exists another problem usually hidden in the mathematical
formalism of quantum theory. Let an external potential be given by
$V(\vec r)$. Transforming Eq. \ref{ws019} into a moving coordinate
system $\vec r\,'$ described by:

\[
        \vec r\,' = \vec r - \vec u \,t
\] 

we get the following equation:

\begin{eqnarray}\label{ws020}
        \left(- \frac{\hbar^2}{2 m} 
        \triangle' + V(\vec r\,' + \vec u \,t) \right) \psi(\vec r\,') = 
        W_{T}\, \psi(\vec r\,')
\end{eqnarray}

Since the variable $t$ in this case is undefined, the relation
contains an element of arbitrariness. It is not conceivable, from
this deduction, how a time--free Schr\"odinger equation could 
be derived: if the time--dependent part of $\psi$ is eliminated,
the procedure is equivalent to a transformation into a moving
coordinate system, but in this case the potential becomes a function
of time. And if the wave equation is integrated in the system at rest,
then the wave functions depend on time. 

From a physical point of view, the reduction of the Schr\"odinger
equation to its time--free form actually eliminates all the
time--dependent interaction and adjustment processes bound to
occur for {\em every} particle wave interacting with a changed
environment. It may well be, that this disregard of development
accounts for the commonly rather static concepts in quantum theory.

\subsection{Uncertainty Relations}

Within the framework of material waves, the arbitrariness of
the Schr\"odinger equation can be referred to the intrinsic
potential of mass oscillations, not accounted for in quantum
theory. And its consequence, as can equally be derived, is
an uncertainty about the exact value of $\vec k$ of the wave
function $\psi(\vec k, \vec r)$. The uncertainty of $\vec k$ 
together with the period of intrinsic potentials $\phi$ is then
sufficient to derive Heisenberg's uncertainty relations 
\cite{HEI26,HEI27}.
To this aim we proceed from the Schr\"odinger equation in the
moving frame of reference Eq. \ref{ws020}. Setting the upper and
lower limits of the potential $V(\vec r' + \vec u t)$ to:

\[
        V_{1} = V(\vec r' + \vec u t_{1}) \quad
        V_{0} = V(\vec r' + \vec u t_{0})
\]

and calculating the correlating $k$--vectors of the wave function
$\psi(\vec r', \vec k)$:

\[
        \psi(\vec r', \vec k) := \psi_{0} e^{i \vec k \vec r'}
\]

we get the following relation:

\begin{eqnarray}\label{ws021}
        \hbar^2 (\vec k_{1}^2 - \vec k_{0}^2) + 2 m (V_{0} - V_{1}) = 0 \\
        \vec k_{1} = \vec k_{1}(\vec r',t) \quad
        \vec k_{0} = \vec k_{0}(\vec r',t) \nonumber
\end{eqnarray}

Evaluating the relation in a one--dimensional model and accounting
for the -- yet unknown -- variation of the potential $V$ by:

\begin{eqnarray}\label{ws022}
        V_{1}(\vec r',t) &=& V(\vec r') + \triangle V(t) \quad
        V_{0}(\vec r',t) = V(\vec r') - \triangle V(t) \nonumber \\
        k_{1}(\vec r',t) &=& k(\vec r') + \triangle k(t) \quad
        k_{0}(\vec r',t) = k(\vec r') - \triangle k(t) \nonumber \\
\end{eqnarray}

the uncertainty $\triangle k$ due to the undefined variable t will be:

\begin{eqnarray}\label{ws023}
        \hbar \triangle k(t) = \frac{m\,\triangle V(t)}{\hbar k}
\end{eqnarray}

If the uncertainty originates from the neglected intrinsic
potential $W_{P}$ (see Eq. \ref{ws011}), the uncertainty of
the applied potentials is described by:

\begin{eqnarray}\label{ws024}
        \phi_{0} = m u^2 = \triangle V
\end{eqnarray}

And therefore the uncertainty of $k$ can be referred to the
value of $k$:

\begin{eqnarray}\label{ws025}
        \hbar \triangle k = \frac{m\,\phi_{0}}{\hbar k} =
        m u = \hbar k
\end{eqnarray}

The uncertainty does not result from wave--features of particles,
as Heisenberg's initial interpretation suggested \cite{HEI27}, 
nor is it an expression of a fundamental physical principle, as the
Copenhagen interpretation would have it \cite{KOP26}. 
It is, on the contrary, an expression of the {\em minimum error}, 
inherent to {\em any} evaluation of the Schr\"odinger equation, 
resulting from the fundamental assumption in quantum theory, i.e.
the interpretation of particles as inertial mass aggregations.

Since the distance between two potential maxima is given by
(the value can deviate in two directions which yields  total
uncertainty of the $x$--coordinate):

\begin{eqnarray}\label{ws026}
        \triangle x = \frac{\lambda}{2} = 2(x(\phi_{0}) - x(0))
\end{eqnarray}

the uncertainty may equally be written as:

\begin{eqnarray}\label{ws027}
        \triangle k \triangle x \ge \frac{k \lambda}{2} \qquad
        \triangle p \triangle x \ge \frac{h}{2}
\end{eqnarray}

Heisenberg actually derived \cite{HEI27} the uncertainty relations
$\triangle p \triangle q = \hbar$. The Fourier transform
inherent to quantum theory
suggests a correction of the constant Eq. \ref{ws027}
by a factor $ (2 \pi)^{-1} $. In this case the deduction yields
{\em exactly} the same result as the canonical 
formulation of quantum theory \cite{COT77}:

\begin{displaymath}
\triangle X_{i} \, \triangle P_{i} \ge \frac{\hbar}{2}
\end{displaymath}

The minimum error due to intrinsic potentials is therefore
equal to the uncertainty of quantum theory,
and the logical relation between Schr\"odinger's equation and
the uncertainty relations is therefore a relation of reason and
consequence: {\em Because} Schr\"odinger's relation is  not precise
by the value of intrinsic potentials, the uncertainty relations
{\em must be} generally valid. This is the second major result
of the present essay:

\begin{itemize}
\item
{\em Because} quantum theory neglects intrinsic characteristics
of particle motion, the Schr\"odinger equation is generally 
not precise by a value described by Heisenberg's uncertainty
relations.
\end{itemize}

The result equally seems to settle the long--standing controversy between
the {\em empirical} and the {\em axiomatic} interpretation of this important 
relation: it is not empirical, since it does not depend on any measurement 
process; but it is equally not a physical principle, because it is due to 
the fundamental assumptions of quantum theory. The solution to the problem
of interpretation is therefore a somewhat wider frame of reference:
although within the principles of quantum theory the relation is an
{\em axiom}, it is nonetheless a {\em result} of fundamental theoretical
shortcomings and not a physical principle.

\subsection{Summary}

In the preceding section we could demonstrate, that quantum theory 
must not necessarily be a comprehensive representation of physical reality.
It was shown, on the contrary, that the principal assumption of quantum
theory, the assumption of inertial mass of particles, is a posteriori
neither justified by the wave features of particles nor by the
fundamental Planck and de Broglie relations. It is but {\em one}
possible theoretical framework and, given the logical problems of
a precise relation between measurements and reality arising from this
concept, not even the best possible framework. 

On an analytical level the establishment of fundamental relations 
of quantum theory within a different framework and their significance
-- that the uncertainty relations can also, and consistently, be
interpreted as an expression of theoretical shortcomings due to
a disregard of intrinsic properties -- served to establish the 
limitations of the current theoretical standard. They also served to
define the logical {\em location} of quantum theory and to determine
the regions of interest, where research is commendable if a theory
{\em outside} the current framework is taken into consideration.

From the viewpoint of theoretical synthesis we showed
that the intrinsic properties of particles and their relations are
a consequence of the experimental evidence of wave--features of 
particles. The de Broglie relations, in a non--relativistic formulation,
provided the essential laws by which these intrinsic variables are
described. They allowed to establish (1) periodic mass--oscillations
within the particle, and (2) the existence of a periodic potential.
These results were sufficient to deduce the fundamental relations of
quantum theory {\em and}, at the same time, to show their inherent 
limitations. Apart from showing the significance of the periodic
potentials -- which will be done in the following sections -- the
first two requirements mentioned in the introduction are therefore
fulfilled. Fundamental results of early quantum theory are reproduced
(Schr\"odinger's relation as well as the uncertainty relations, while 
single experiments will be treated in the following sections), and the 
intrinsic wave features of particles were shown to be
of {\em physical} significance, which strongly supports Einstein's
view in the EPR paradox.


\section{Classical Electrodynamics}\label{sec_ed}

Classical electrodynamics is considered one of the most
refined and powerful physical theories at hand. It served,
during the last hundred years, ever again as a model for 
physical formalization, and its implications led to quite a
few extensions of physical knowledge, not least Albert Einstein's
theory of Special Relativity \cite{EIN05A}. A great deal of
the fascination of classical electrodynamics results from its
mathematical conciseness and brevity. The possibility to 
describe and account for most of the phenomena concerning
interactions of charge by a set of basically four linear 
differential equations, the {\em Maxwell equations} \cite{SOM64},
must either seem a lucky coincidence or an expression of
a deeply rooted analogy in physical reality.

That classical electrodynamics cannot be referred to mechanical
conceptions, has been proved during the scientific controversy
about the existence of the ''{\em Aether}''.
This proposed medium of electromagnetic fields was never detected,
its existence, furthermore, would contradict the experimental results of
Michelson and Morley \cite{MIC87}. From the theoretical side, the
controversy was settled by Einstein's theory of relativity. 
Since the correctness of Maxwell's equations is established
beyond doubt, it remains an epistemological problem, why no
attempt to refer them to fundamental physical axioms has 
succeeded so far \cite{JAC84}. Summarizing the current situation
in two statements: (1) Maxwell's equations are correct, and 
(2) they cannot be referred to mechanical axioms, there exist
equally two logical conclusions to the epistemological problem:
(1') Maxwell's equations are themselves of an axiomatic nature,
and (2') The axioms, on which they are based, have not yet been
detected.

Currently, the established opinion is (1'). Basically, one could
leave it at this state and employ the relations as correct
algorithms. From the viewpoint of macro physics, the procedure
is justified by the wealth of experimental verifications at hand.
The same does not apply, though, to an application of 
electrodynamics in micro physical models. Since the exact origin
of the relations is obscure, it cannot be guaranteed, that they
do not, in some way, preclude characteristics of physical 
systems, which contradict other assumptions. Obviously, there
exist only two ways out of this logical dilemma: either it can
be proved, that quantum theory is logically independent of
electrodynamics -- a proof, which never has been given --, or
it can be established that electrodynamics itself is based on
micro physical axioms: in this case the current procedure in
QED \cite{SCH94} is logically insufficient. The logical
conclusion to our epistemological problem would then be:
(2'') The Maxwell relations are an expression of intrinsic
characteristics of particles.

\subsection{The Maxwell equations}

Based on the intrinsic wave characteristics of free particles,
developed in the previous section, the verification of (2'')
does not present an impossible problem. The Maxwell equations,
in material wave theory, are an expression of longitudinal 
variations of intrinsic variables of particles in motion.
From the wave equation for longitudinal momentum $\vec p$:

\begin{eqnarray}\label{ed001}
\nabla^2\,\vec p - \frac{1}{u^2} \frac{\partial^2\,\vec p}{\partial t^2}
= 0 \quad \vec p = \rho \,\vec u \quad \vec u = \mbox{const}
\end{eqnarray}

And the equation of continuity for momentum $\vec p$  and density
$\rho$:

\begin{eqnarray} \label{ed002}
\nabla\,\vec p + \frac{\partial\,\rho}{\partial t} = 0
\end{eqnarray}

it follows within a micro volume, where particle velocity 
$\vec u$ is constant:

\begin{eqnarray}\label{ed003}
\nabla^2\,\vec p & = & - \nabla \,\frac{\partial\,\rho}{\partial t}
- \nabla \times (\nabla \times \vec p) \nonumber \\
\frac{1}{u^2} \frac{\partial^2\,\vec p}{\partial t^2} & = &
\frac{\partial}{\partial t}  \left(
\frac{\bar{\sigma}}{u^2} \frac{1}{\bar{\sigma}}
\frac{\partial\,\vec p}{\partial t} \right) 
\end{eqnarray}

The parameter $\bar{\sigma}$ is a dimensional constant
employed to make mechanical units compatible with electromagnetic
variables, its dimension is equal to density of charge. 

Electric fields $\vec E$ and magnetic fields $\vec B$ are 
now defined in terms of momentum $\vec p$ and the potential
$\phi (\vec r,t)$ due to mass oscillations (see the previous
section):

\begin{eqnarray}
\vec E &:=& - \nabla\,\frac{1}{\bar{\sigma}} \phi(\vec r,t) + 
\frac{1}{\bar{\sigma}} 
\frac{\partial\,\vec p}{\partial t} \label{ed004}\\
\vec B &:=& - \frac{1}{\bar{\sigma}}
\nabla \times \vec p \label{ed005}
\end{eqnarray}

And substituting Eq. \ref{ed003} into \ref{ed001} we get, with
the help of \ref{ed004} and \ref{ed005}:

\begin{eqnarray}\label{ed006}
\frac{\partial}{\partial t} \, \nabla
\left( \frac{1}{u^2}\, \phi + \rho \right) +
\bar{\sigma}\left( 
\frac{1}{u^2} \frac{\partial \vec E}{\partial t} - 
\nabla \times \vec B \right) = 0
\end{eqnarray}

And since the total intrinsic potential is a constant:

\begin{eqnarray} \label{ed007}
\phi(\vec r,t) + \rho(\vec r,t)\,u^2 = \phi_{0} = \mbox{const}
\end{eqnarray}

The following relation holds in full generality within 
every micro volume:

\begin{eqnarray} \label{ed008}
\frac{1}{u^2} \frac{\partial \vec E}{\partial t} =
\nabla \times \vec B
\end{eqnarray}

Rotation of Eq. \ref{ed004} yields, but for a constant,
Maxwell's first equation:

\begin{eqnarray}\label{ed009}
\nabla \times \vec E = 
\frac{1}{\bar{\sigma}} 
\frac{\partial}{\partial t} \nabla \times \vec p =
- \frac{\partial \vec B}{\partial t}
\end{eqnarray}

To deduce the inhomogeneous equation we additionally have 
to employ the source equations of electrodynamic theory and 
the definition of $\vec H$ fields:

\begin{eqnarray}
\nabla ( \epsilon \vec E ) = \nabla \vec D = \sigma 
\quad \nabla \vec B = 0 \\
\nabla J(\vec r,t) = - \frac{\partial \sigma}{\partial t} \label{ed010}\\
\vec H = \mu^{-1} \,\vec B
\end{eqnarray}

It is assumed that the material constants $\epsilon$ and $\mu$
also remain constant within the micro volume. Substituting into
Eq. \ref{ed008} and computing the source of \ref{ed010}, the
sum yields:

\begin{eqnarray}\label{ed011}
\nabla \left(
- \frac{1}{u^2} \frac{\partial \mu^{-1} \vec E}{\partial t} + 
\frac{\partial \vec D}{\partial t} + \vec J \right) = 0
\end{eqnarray}

And if the constant of integration is zero, the result will
be Maxwell's inhomogeneous equation:

\begin{eqnarray}
\vec J + \frac{\partial \vec D}{\partial t} = \nabla \times \vec H
\end{eqnarray}

Finally, by computing the time differentials of Eqs. \ref{ed008}
and \ref{ed009}, the wave equations for electromagnetic fields
in a vacuum ($\sigma = 0$) can be deduced:

\begin{eqnarray}
\nabla^2 \vec E - \frac{1}{u^2} 
\frac{\partial^2 \vec E}{\partial t^2} = 0 \label{ed012}\\
\nabla^2 \vec B - \frac{1}{u^2} 
\frac{\partial^2 \vec B}{\partial t^2} = 0 \label{ed013}
\end{eqnarray}

The results indicate, that electrodynamics and quantum theory
do have a common root, the {\em intrinsic} properties of moving
particles. From the viewpoint of material waves they are therefore
different aspects of the same physical phenomenon, the wave
features of mass, which is the third major result of this
essay:

\begin{itemize}
\item
Electrodynamics and wave mechanics are theoretical concepts
based on the same physical phenomena, the intrinsic features of
moving particles.
\end{itemize}

Logically speaking, the result means, that a method, which employs
quantum theory to calculate physical variables by way of observables,
and which identifies these observables -- measured variables --
as field variables in electrodynamics cannot be correct. 

\subsection{Energy relations}

The energy principle of material waves can be identified as the classical
Lorentz standard by a combination of the continuity equation and the 
definition of electric fields including intrinsic potentials. Since:

\begin{eqnarray}\label{ed014}
        \nabla {\vec p} + 
        \frac{\partial \rho}{\partial t} = 0
        \qquad
        \bar \sigma {\vec E} = 
        - \nabla \phi + \frac{\partial {\vec p}}{\partial t}
\end{eqnarray}

the wave equation for linear motion 
$ \nabla \times \nabla \times {\vec p} = 0 $ leads to:

\begin{eqnarray}
        \nabla (\nabla {\vec p}) + \frac{\partial}{\partial t} \nabla \rho
        = \frac{1}{u^2} \frac{\partial^2}{\partial t^2} {\vec p} 
        + \frac{\partial}{\partial t} \nabla \rho
        \\ \Rightarrow \qquad 
        \frac{\partial {\vec p}}{\partial t} + 
        u^2 \nabla \rho = const =: 0
        \label{ed015}
\end{eqnarray}

And for uniform motion $ \bar \sigma {\vec E} = 0 $ the
potential can be related to momentum $ {\vec p} $ by:

\begin{displaymath}
        \frac{1}{u^2} \frac{\partial \phi}{\partial t} -
        \nabla {\vec p} = 0
\end{displaymath}

Comparing with the expression in classical electrodynamics the vector
potential $\vec A$, in classical electrodynamics an abstract conception,
is proportional to momentum of particle motion:

\begin{eqnarray}
        &\nabla {\vec A}& + \frac{1}{c} 
        \frac{\partial \phi}{\partial t} = 0
        \qquad
        - \nabla {\vec p} + \frac{1}{u^2} \frac{\partial \phi}
        {\partial t} = 0 \nonumber\\
        &{u = c \atop \Rightarrow}& \qquad 
        {\vec A} = - c \, {\vec p} =: \frac{1}{\alpha} 
        \,{\vec p}\label{ed016}
\end{eqnarray}

\subsection{Photons}

The wave--features of photons, or the representation of electromagnetic
fields by wave--type field vectors, have long been the only aspect
in consideration. The picture changed only after experimental research
was directed towards the fundamental problems of interaction between
matter and electromagnetic fields. Due to Planck's result on black--body
radiation and Einstein's application of Planck's formula to the problem
of photo--electricity \cite{EIN05B}, the energy of photons gained special
attention, since it was (1) proportional to frequency and not, as field
theory proposed, intensity, and (2) the energy quantum $\hbar \omega$
could no longer be distributed locally, because the energy changes in
interactions required an exchange of a full number $n \in N$ of energy
quanta. In Einstein's view the interaction suggested that energy of
photons be distributed in single points within the electromagnetic fields
of radiation.

According to the Copenhagen interpretation of quantum theory \cite{KOP26} 
the two theoretical aspects -- wave--features and energy quanta -- describe
complementary facts of a complex physical reality not adaptable to
any single  theoretical framework. This interpretation, which focusses
on the central notion of wave--particle--duality, states that all
the qualities like polarization and interference, typical for a
wave--type theory, have to be treated by electrodynamics, while
energy and interaction phenomena must be seen from the viewpoint of
quantum theory. The logical justification for retaining the duality,
which is in itself a logical contradiction if wave--features and
particle features were to prevail {\em simultaneously}, was seen by
Jordan \cite{JOR36} and Heisenberg \cite{HEI41} in the impossibility
of experiments -- i.e. physical processes -- where both aspects
{\em must} simultaneously exist.

This justification is not undisputed. Renninger showed by a rather 
simple thought--experiment \cite{REN53}, that an interference measurement
involving two separate light--paths does not allow for the
probability--interpretation of electromagnetic field vectors, since the result
of the measurement can be altered in a deterministic manner by 
the insertion of optical devices into a single light--path. The
analysis of the experiment leads to the conclusion, that the
wave--features {\em must have} physical significance, which again seems
to confirm the EPR dilemma already quoted \cite{EPR35}. Based on this
experiment Renninger deduced the following statement about the
features of photons (translation by the author, emphasizing by
Renninger):

{\em Every light--quantum consists of an energy--particle, which is guided
by a wave without energy}

The important result, from our point of view, is the physical significance
of wave--features. The energy particle, which Renninger refers to, is a
direct result of Einstein's interpretation of photo--electricity, which
equally determines the interpretation of the guiding wave as energy--free.
From the viewpoint of physical concepts, an energy--free {\em physical}
wave is not all too convincing. If a region of space is altered by a
propagating wave, thus the obvious objection, then it cannot be without
a physical entity moving. But since all physical field--variables are, in some
way, related to energy, a physical entity propagating without any
propagation of energy seems contradictory. 

On thorougher analysis the necessity to divide between two separate
physical entities to account for energy as well as electromagnetic
features of photons depends on the interpretation of the interaction
process itself. Since (1) the quantum of energy is fully transferred, 
and (2) the transfer interval is well below the level of measurements,
the energy must be contained in an insignificantly small volume. The
alternative being, as Renninger pointed out, that the wave will be
contracted with a velocity exceeding $c$ during every interaction 
process. 

It is this interpretation of the interaction process, which will be
changed by material wave theory. To understand the decisive modification,
we present here -- and without proof -- a result on the interactions of
electrons and photons, which will be derived later on. 

Assuming that the interaction interval equals \mbox{$n\,\tau, n \in N$}, and
setting $n$ for convenience equal to 1, an intrinsic model of particles
and fields will have to consider energy density at a specific point
{\em within} the particle. If total energy density is constant and
$\phi_{0}$, then the time--derivative of energy density equals:

\begin{eqnarray}\label{ed017}
\frac{d \phi}{d t} &=& \frac{\phi_{0}}{\tau} =
\frac{\rho u^2}{\tau} = \frac{1}{V_{ph}}\, h\nu^2 \nonumber\\
\frac{d E_{ph}}{d t} &=& V_{ph} \frac{d \phi_{0}}{d t} = h \nu^2 \\ 
E_{ph} &=& \int_{0}^{\tau} dt \frac{d E_{ph}}{d t} = h \nu \nonumber 
\end{eqnarray}

In this way the traditional features of photons are retained, while
the meaning of the term {\em energy quantum} is modified: the energy
quantum {\em transferred} in an interaction process equals $h \nu$, 
regardless of the actual volume of the photon, and equally regardless
of the exact duration of the interaction process. As long as the total
density of energy $\phi_{0}$ is constant, the relation holds for every
single point of the underlying field and therefore for the whole region
of the physical wave corresponding to a photon. 

Now let us consider a case, where the initial photon is, by some optical
device, split into two. If one of the partial volumes -- a half photon
so to say -- then interacts with matter, the process is described by:

\begin{eqnarray}\label{ed018}
\frac{d \phi}{d t} &=&
\frac{\rho u^2}{\tau} = \frac{1}{2 V_{ph}'}\, h\nu^2 \nonumber\\
\frac{d E_{ph}'}{d t} &=& V_{ph}' \frac{d \phi_{0}}{d t} = \frac{1}{2}
h \nu^2 = \frac{1}{2} \frac{d E_{ph}}{d t} \\ 
E_{ph}' &=& \int_{0}^{\tau} dt \frac{d E_{ph}'}{d t} = \frac{1}{2}
h \nu = \frac{1}{2} E_{ph} 
\nonumber 
\end{eqnarray}

The energy density transferred remains constant, but not total energy.
The process therefore {\em seems} to be a confirmation of
the principle, that all interactions occur in discrete energy quanta
only, if discrete quantities of mass or charge are considered. Only,
in short, if the general outlook on processes is a mechanical one.
It is no contradiction with Einstein's calculation on photo--electricity,
to interpret it as an effect of {\em energy densities} and transfer
rates. It only has to be considered, that the
total volume affected by interactions will depend on the total volume
of the -- fractional -- photon.

From this result it can be concluded, that the term {\em energy quantum}
is not appropriate for the physical process of interaction: since the
total energy transferred depends on (1) the frequency of radiation, and
(2) the volume of the photon, total energy transferred is the product of
energy density -- proportional to frequency -- and volume -- which may be
related to intensity.  In an {\em intrinsic} formalism of interaction,
furthermore, the total quantities transferred are only secondary, the
more so, since electrodynamics already is a theory of intrinsic wave
features, as deduced above.

In material wave theory photons are distributions of mass and
energy, and the guiding wave possesses the same dimensions as the
energy distribution, while the guiding wave as well as mass movement
contribute alike to total energy of the photon. To derive the
essential features of photons we use the fundamental wave equation
of intrinsic particle momentum. The exact dimensions as well as the
problems of photon boundaries are shown to be of no effect for the
results achieved. 

Denoting an, initially, abstract momentum of a photon by $ { \vec p} $,
we proceed from the wave equation in a vacuum given by:

\begin{eqnarray}\label{ed019}
        \nabla^2\, { \vec p} - \frac{1}{c^2}\, 
        \frac{\partial^2\, { \vec p}}{\partial \,t^2} = 0
\end{eqnarray}

With the identity for the Laplace operator and the general relation
between momentum $ { \vec p} $ and vector potential $ { \vec A} $ 
from Eq. (~\ref{ed016}):

\begin{eqnarray}
        \nabla^2\, { \vec V} &=& \nabla (\nabla \,{ \vec V}) - 
        \nabla \times (\nabla \times { \vec V}) \nonumber\\
        { \vec p} &=& \alpha { \vec A} \nonumber \\ 
        \frac{1}{c} \frac{\partial \,{ \vec p}}{\partial\, t} &=&
        \frac{\alpha}{c} \frac{\partial \,{ \vec A}}{\partial\, t} \qquad
        \nabla \times { \vec p} = \alpha\, \nabla \times { \vec A}
        \label{ed020}
\end{eqnarray}

the wave equation may be rewritten as:

\begin{equation}\label{ed021}
        \nabla ( \nabla \,{ \vec p} ) - \alpha\, \nabla \times { \vec B}
        - \frac{\alpha}{c} \left(\frac{1}{c} 
        \frac{\partial^2 \,{ \vec A}}{\partial\, t^2}
        \right) = 0
\end{equation}

Using the solution for the homogeneous Maxwell relations
(the following deduction employs, for convenience, Gaussian units):

\begin{equation}\label{ed022}
        \frac{1}{c} \frac{\partial \,{ \vec A}}{\partial\, t}
         = - \nabla\, \phi ({ \vec r}, t) - { \vec E}({ \vec r}, t)
\end{equation}

the wave equation yields the following result:

\begin{eqnarray}
        &\nabla& \left( \nabla \,{ \vec p} + 
        \frac{\alpha}{c} 
        \frac{\partial\, \phi}{\partial \,t} \right)
        - \alpha\, \underbrace{
        \left( \nabla \times { \vec B} - 
        \frac{1}{c} \frac{\partial \,{ \vec E}}{\partial\, t} \right) }
        _{= 0}
        = \nonumber\\ &=& \nabla \left( \nabla \,{ \vec p} + 
        \frac{\alpha}{c} \frac{\partial\, \phi}{\partial \,t} \right) = 0
        \label{ed023}
\end{eqnarray}

We choose the specific solution by setting the term in brackets equal
to zero and get therefore:

\begin{equation}\label{ed024}
        \nabla \,{ \vec p} + 
        \frac{\alpha}{c} \frac{\partial\, \phi}{\partial \,t} = 0
\end{equation}

The equation can be solved by any wave packet consisting of
plain waves and complementary potentials given by:

\begin{eqnarray}
        { \vec p} &=& \sum_{i} p_{i}^{0} { \vec e}^{k_{i}} 
        \sin^2 ({ \vec k}_{i} { \vec r} - \omega_{i} t)
        \nonumber\\
        \phi &=& \sum_{i} \phi_{i}^{0} 
        \cos^2 ({ \vec k}_{i} { \vec r} - \omega_{i} t)\label{ed025}
\end{eqnarray}

For a component $ { \vec p}_{i}, \phi_{i} $ of the wave packet  
the amplitudes must comply with:

\begin{equation}\label{ed026}
        \left(p_{i}^{0} k_{i} + \frac{\alpha\, \omega_{i}}{c} \,
        \phi_{i}^{0}\right) = 0
\end{equation}

Setting $ \alpha $ equal to $ - 1/c $ and using the 
amplitudes of material waves as well as vacuum dispersion
we recover Einstein's energy relation ~\cite{EIN06}:

\begin{eqnarray}
        p_{i}^{0} = \rho_{ph, i}^{0}\, c \qquad c = \frac{\omega_{i}}{k_{i}}
        \nonumber \\
        \phi_{i}^{0} = \rho_{ph, i}^{0} \,\frac{c^2 \,c k_{i}}{\omega_{i}}
        = \rho_{ph, i}^{0} \,c^2 \label{ed027}
\end{eqnarray}

We define now a {\em photon} as any component of the wave
packet. More specifically, we describe a photon by a plane material
wave and its complementary electromagnetic potential.

\begin{eqnarray}
        { \vec p} \,:= \, { \vec p}_{i} \,=\, \rho_{ph}^{0}\, c\, 
        { \vec e}^{k}
        \sin^2 ({ \vec k} { \vec r} - \omega t) \nonumber\\
        \phi \, := \, \phi_{i} \,=\, \rho_{ph}^{0}\, c^2\, 
        \cos^2 ({ \vec k} { \vec r} - \omega t)\label{ed028}
\end{eqnarray}

The {\em energy density} at a specific location of the photon
is denoted by an electromagnetic potential and a potential
due to the vector $ {\vec p} $. If we identify $ {\vec p} $ as
a longitudinal material wave, its energy density at any arbitrary
point, the {\em total potential},
is double its value in classical mechanics.

\begin{eqnarray}
        \phi_{k}({ \vec r}, t) = { \vec p}\, c \,{ \vec e}^{k}
        =  \rho_{ph}^{0}\, c^2 \,
        \sin^2 ({ \vec k} { \vec r} - \omega t) \label{ed030}\\
        \phi_{e}({\vec r}, t) = \phi ({\vec r}, t)
        =  \rho_{ph}^{0}\, c^2 \, \cos^2 
        ({\vec k} {\vec r} - \omega t) \label{ed031}
\end{eqnarray}

Then the material aspect of a photon in motion is that of a
material wave, while, due to additional electromagnetic characteristics,
total energy density is constant and does not
depend on location or time:

\begin{equation}\label{ed032}
        \phi_{ph}^{0} \, := \phi_{k} + \phi_{e} =  
        \rho_{ph}^{0}\, c^2
        \end{equation}

The electromagnetic potential must correlate with electromagnetic
field vectors. Using electromagnetic potentials of a vacuum:

\begin{equation}\label{ed033}
        \phi_{e} = \frac{1}{8 \pi} 
        \left( { \vec E}^2 + { \vec B}^2 \right)
\end{equation}

and assuming for consistency with classical electrodynamics, that 
electromagnetic fields will be of transversal polarization,
the transversal fields of a photon will be given by:

\begin{eqnarray}
        \frac{1}{8 \pi} \left( { \vec E}^2 + { \vec B}^2 \right) 
                &=& \rho_{ph}^{0}\, c^2 \, 
                \cos^2 ({ \vec k} { \vec r} - \omega t)
        \nonumber\\
                { \vec E} ({ \vec r}, t) &=& { \vec E}_{0}\, 
                \cos ({ \vec k} { \vec r} - \omega t) 
        \\      
                { \vec B} ({ \vec r}, t) &=& { \vec B}_{0}\, 
                \cos ({ \vec k} { \vec r} - \omega t)
        \nonumber
\end{eqnarray}
\begin{eqnarray}
                { \vec E}_{0} &=& c\, 
                \sqrt{4 \,\pi\,\rho_{ph}^{0} }\, { \vec e}^{t}
        \nonumber\\
                { \vec B}_{0} &=& c\, 
                \sqrt{4 \,\pi\,\rho_{ph}^{0} } \,
                ({ \bf e}^{k} \times { \vec e}^{t}) 
        \\
                { \vec e}^{k}  { \vec e}^{t} &=& 0 
        \nonumber
\end{eqnarray}

Transversal electromagnetic fields of a wave packet are 
therefore described by:

\begin{eqnarray}
        { \vec E} &=&  c\, \sum_{i} 
                \sqrt{4 \,\pi\,\rho_{ph, i}^{0} }\, { \vec e}_{i}^{t}
                \cdot\nonumber\\
                &\cdot&\left(e^{i ({ \vec k}_{i} { \vec r} - \omega_{i} t)} +
                e^{- i ({ \vec k}_{i} { \vec r} - \omega_{i} t)} \right)
        \nonumber\\
                { \vec B} &=&  c\, \sum_{i} 
                \sqrt{4 \,\pi\,\rho_{ph, i}^{0} }\,
                ({ \vec e}^{k} \times { \vec e}^{t})\cdot
                \nonumber\\
                &\cdot&
                \left(e^{i ({ \vec k}_{i} { \vec r} - 
                \omega_{i} t)} +
                e^{- i ({ \vec k}_{i} { \vec r} - 
                \omega_{i} t)} \right)
\end{eqnarray}

The units of electromagnetic fields will be treated on a more 
fundamental basis in the following sections, presently they
only signify a constant value of some suitable dimension.

\subsection{Special Relativity}

In the section on electrodynamics of his famous publication
Einstein \cite{EIN05A} referred the universal validity of 
Maxwell's equations to their Lorentz--invariance. Based on
the theoretical framework of intrinsic features of variables,
it is tempting to relate the Lorentz--invariance of Maxwell's
equations to the invariance of the wave equations under a
Lorentz transformation. If it can be established, that the Lorentz
transformation is also the transformation leaving intrinsic
features of particles intact, then the kinematic section of
the theory of relativity can also be seen as a necessary
modification of mechanical concepts to account for electromagnetic
phenomena. 

As the Maxwell relations are not only valid for intrinsic features
of photons but also electrons -- which derives from their independence
of particle velocity u -- we proceed from the wave equation for
the intrinsic momentum $\vec p$ of a particle with arbitrary velocity
$u = u_{x}$ in a moving frame of reference S' with 
$\vec V = V \vec e_{x}$ relative to a system S at rest. The 
wave equation in S' then will be:

\begin{eqnarray}\label{ed035}
\triangle' p_{x}' - \frac{1}{u_{x}'^2} 
\frac{\partial^2 p_{x}'}{\partial t'^2} = 0 \qquad
p_{x}' = \rho' u_{x}'
\end{eqnarray}

With the standard Lorentz transformation:

\begin{eqnarray}\label{ed036}
x'^{\mu} = \Lambda^{\mu}_{\nu} x^{\nu} \qquad
\Lambda^{\mu}_{\nu} = \left(
\begin{array}{rrrr}
\gamma & - \beta \gamma & 0 & 0 \\
- \beta \gamma & \gamma & 0 & 0 \\
0 & 0 & 1 & 0 \\
0 & 0 & 0 & 1 
\end{array}
\right)
\end{eqnarray}

The differentials in S' are given by:

\begin{eqnarray}\label{ed037}
\triangle' = 
\left(\frac{\partial x}{\partial x'}  \right)^2 \triangle =
(1 - \beta^2) \triangle        \\
\frac{\partial^2}{\partial t'^2} = 
\left(\frac{\partial t}{\partial t'}  \right)^2
\frac{\partial^2}{\partial t^2}  = (1 - \beta^2) 
\frac{\partial^2}{\partial t^2} \nonumber
\end{eqnarray}

Density of mass and velocity in S' are described by 
(transformation of velocities according to the 
Lorentz transformation, $\alpha$ presently undefined):

\begin{eqnarray}\label{ed038}
\rho' = \alpha \cdot \rho \qquad
u_{x}' = \frac{u_{x} - V}{1 - \frac{u_{x} V}{c^2}}
\end{eqnarray}

Then the transformation of the wave equations yields two
wave equations for momentum $p_{x}$ and density $\rho$:

\begin{eqnarray}\label{ed039}
\triangle p_{x} - \frac{1}{u_{x}'^2} 
\frac{\partial^2 p_{x}}{\partial t^2} = 0 \nonumber \\
\triangle \rho - \frac{1}{u_{x}'^2} 
\frac{\partial^2 \rho}{\partial t^2} = 0
\end{eqnarray}

Energy measurements in the system S at rest must be measurements
of the phase--velocity $c_{ph}$ of the particle waves. Since,
according to the transformation, the measurement in S yields:

\begin{eqnarray}\label{ed040} 
c_{ph}^2 (S) = u_{x}'^2
\end{eqnarray}

the total potential $\phi_{0}$, measured in S will be:

\begin{eqnarray}\label{ed041}
\frac{\phi_{0}(S)}{\rho_{S}} = u_{x}'^2  \quad \Rightarrow \quad
\phi_{0} (S) = \frac{\rho_{S}}{\rho_{S'}} \,\phi_{0} (S')
\end{eqnarray}

And if density $\rho_{S'}$ transforms according to a relativistic
mass--effect with:

\begin{eqnarray}\label{ed042}
\rho_{S'} = \gamma \, \rho_{S}
\end{eqnarray}

then the total intrinsic potential measured in the system S' will
be higher than the potential measured in S:

\begin{eqnarray}\label{ed043}
\phi_{0} (S') = \gamma \phi_{0} (S)
\end{eqnarray}

But the total intrinsic energy of a particle with volume $V_{P}$
will not be affected, since the x--coordinate will be, again 
according to the Lorentz transformation, contracted:

\begin{eqnarray}\label{ed044}
V_{P} (S') &=& \frac{1}{\gamma}\,V_{P} (S) \nonumber \\
E_{0} (S) &=& \phi_{0} (S) V_{P} (S) = 
\frac{1}{\gamma}\,\phi_{0}(S') \, V_{P} (S) = \nonumber \\
&=& \phi_{0} (S') V_{P} (S') = E_{0} (S')
\end{eqnarray}

The interpretation of this result exhibits some interesting
features of the Lorentz transformation applied to electrodynamic
theory. It leaves the wave equations and their energy relations
intact if, and only if energy {\em quantities} are evaluated,
i.e. only in an integral evaluation of particle properties.
In this case it seems therefore justified, to consider 
Special Relativity as a necessary adaptation of kinematic
variables to the theory of electrodynamics. The same does not
hold, though, for the intrinsic potentials of particle motion.
Any transformation into a moving reference frame in this case
leads to an increase of the total potential $\phi_{0}$, in the
case of photons ($\gamma \rightarrow \infty$) the intrinsic
potential then will be infinite. The essential result of this
analysis of Special Relativity may be expressed by:

\begin{itemize}
\item
The wave equations of intrinsic particle structures are
Lorentz invariant only for an integral evaluation of energy.
The total intrinsic potentials increase for every transformation
into a moving frame of reference.
\end{itemize}

\subsection{Summary}

That Maxwell's relations can be referred to intrinsic 
characteristics of free particles, proves two assumptions:
(1) The relations are themselves not of an axiomatic nature,
contrary to current opinions they can be referred to a 
sublayer of still more fundamental axioms. (2) Quantum theory
and electrodynamics are not logically independent, since both
concepts are a formalization of wave features of particles.

The features of photons in material wave theory equally seem to
fit into the framework of classical electrodynamics: electromagnetic
field vectors are transversal, while the longitudinal component
is purely kinetic. Both aspects together provide a model of
photons, which is simultaneously an extension of electrodynamics
(since the variables of motion exist in addition to the 
electromagnetic field--vectors), and an extension of quantum
theory (since the electromagnetic fields are not originally
included in the kinetic model of quantum theory).   

Relativistic electrodynamics only provides
Lorentz invariant formulations of intrinsic particle
features, if energy quantities are evaluated. Since the
intrinsic potentials depend on the reference frame of evaluation
and increase by a transformation into a moving coordinate
system, the theory suggests that the infinity problems,
inherent to relativistic quantum fields, may be related to
the logical structure of QED. 


\section{Electron photon interactions}

\subsection{Magnetic properties of particles}

While the definition of magnetic fields Eq. \ref{ed005}
allows to deduce the Maxwell equations and accounts for
intrinsic properties of photons, it is incorrect, by a
factor 2, if magnetostatic interactions of electrons 
are considered. The definitions, describing the relations between
the kinetic variables $ (\rho, \vec u)$ and the magnetic field
$(\vec B)$, of photons and electrons correctly are written:

\begin{eqnarray}\label{ep000}
\vec B_{ph} (\vec r,t) = - \frac{1}{\bar \sigma} \nabla
\times \vec p_{ph} (\vec r,t) = \vec B_{intr} \nonumber \\
\vec B_{el} (\vec r,t) = - \frac{1}{2 \bar \sigma} \nabla
\times \vec p_{el} (\vec r,t) = \vec B_{ext}
\end{eqnarray}

The reason for this difference, which bears a superficial resemblance
to the distinction between bosons \mbox{(s = $\pm 1$)} and fermions 
(s = $\pm \frac{1}{2}$), is that the two relations refer to different
magnetic fields. On thorougher analysis, two different cases have to
be considered: (1) The intrinsic magnetic fields, deriving from the
kinetic properties of motion and oscillation, and (2) the external
magnetic fields due to curvilinear motion of charge. While for photons
the intrinsic electromagnetic fields have been known and dealt with
in the framework of classical electrodynamics, for electrons these
intrinsic fields can only be derived within material wave theory but
not in quantum theory. And while the Maxwell relations as an expression
of intrinsic features are valid for both types of particles, the
magnetostatic properties of charge apply only for electrons. The
correction arises therefore from the treatment of different 
physical processes, and is easily understood as a reminiscence to
the historical origins of the separate theories.

It is tempting, at this stage, to refer the difference between
fermions and bosons to the same historical origins, adopted for the
concept of spin without further reflection. That the spin of
particles can indeed be referred to magnetic fields, will
be shown in section \ref{ep_spin}.

\subsection{Static interactions} 

\subsubsection{Magnetostatics}

To account for magnetostatic interactions the definition of 
magnetic fields therefore has to be corrected.

In {\em homogeneous} magnetic fields 
$ { \vec B} = B_{0} { \vec e}^{z} $ electron 
motion can be treated  based on the changed definition of 
external magnetic fields Eq. (\ref{ep000}):

\begin{eqnarray}\label{ep001}
        { \vec B} ({ \vec r}, t) &=&
        - \frac{1}{2 \bar \sigma} \nabla \times 
        \rho ({ \vec r}, t)\,{ \vec u} ({ \vec r}, t) \\
        B_{0} { \vec e}^        {z}  &=&
        - \frac{\rho}{2 \bar \sigma} 
        \nabla \times { \vec u} + \frac{ \vec u}{2 \bar \sigma} 
        \times \nabla\,\rho = 
        \frac{\rho}{2 \bar \sigma} \nabla \times \,{ \vec u} \nonumber
\end{eqnarray}

The sources of $ \rho $ vanish, since:

\begin{equation} \label{ep002}
        { \vec k} = \frac{m}{\hbar}\,{ \vec u} \quad \Rightarrow \quad
        \nabla \rho = f'({ \vec r}, t)\,{ \vec u} \nonumber
\end{equation}

The solution is equivalent to the traditional one:

\begin{equation}\label{ep003}
        { \vec u} = u_{0}\, { \vec e}^{z} + 
        B_{0} \frac{e}{m}\,
        { \vec e}^{z} \times { \vec r}
\end{equation} 

\subsubsection{Lorentz forces}

Circular rotation in homogeneous magnetic fields is
not subject to quantization, because the local component
of de Broglie's relation vanishes. Particle density
$\rho(\vec r,t)$ in any state of rotation is given by:

\begin{eqnarray}\label{ep110}
\rho (\vec r,t) = \rho_{0} \sin^2 (\vec k \vec r - \omega t)
\quad
\vec k \vec r = \frac{m}{\hbar} (\vec \omega \times \vec r)
\cdot \vec r = 0 \nonumber \\
\rho = \rho (t) = \rho_{0} \sin^2 (\omega t)
\end{eqnarray}

Then the relation between magnetic field $\vec B$ and 
angular velocity $\omega$ will be, with Eq. \ref{ep000}:

\begin{eqnarray}\label{ep111}
\vec B = - \frac{\rho(t)}{2 \bar \sigma} 
\nabla \times (\vec \omega \times \vec r)
\end{eqnarray}

And since $\vec \omega \perp \vec r$ it follows generally:

\begin{eqnarray}\label{ep112}
\vec B = \frac{\rho}{\bar \sigma} \,\vec \omega
\end{eqnarray}

Considering the classical formulation for Lorentz forces
of charge in a magnetic field $ \vec B$, we may write:

\begin{eqnarray}\label{ep113}
\vec F_{L} = \sigma (\vec u \times \vec B) =
- \rho\,\frac{\sigma}{\bar \sigma}\, \vec \omega \times
( \vec \omega \times \vec r) = - \rho\,\frac{\sigma}{\bar \sigma}
\omega^2 \vec r
\end{eqnarray}

Which reduces in the average $\sigma = \bar \sigma$ to:

\begin{eqnarray}\label{ep114}
\vec F_{L} = - \rho \omega^2 \vec r
\end{eqnarray}

Additionally, it has to be considered that rotation of mass
in the rotating coordinate system gives rise to centrifugal
forces described by:

\begin{eqnarray}\label{ep115}
\vec F_{C} = \rho \omega^2 \vec r
\end{eqnarray}

Then the rotation of charge is, in the rotating coordinate system,
free of forces, it is therefore its {\em inertial} state of
motion in a magnetic field $\vec B$:

\begin{eqnarray}\label{ep116}
\vec F = \vec F_{L} + \vec F_{C} = 0
\end{eqnarray}

But as centrifugal forces are inertial forces arising from the
curvilinear path of motion, the same must apply to Lorentz forces:
a magnetic field determines rotation of charge, which experiences
centrifugal and Lorentz forces. Which means, evidently, that
Lorentz forces must be inertial forces.

\subsubsection{Current}

The framework developed so far has some interesting
consequences in electrostatic problems. Defining a vector field 
$ { \vec j}({ \vec r}, t) $ by:

\begin{equation}\label{ep004}
        \nabla \,{ \vec j}({ \vec r}, t) \equiv
        - \frac{\partial \,\sigma}{\partial \,t}
        \qquad \sigma = \frac{e}{m} \,\rho
\end{equation}

and using the equation of continuity, which is equally valid for electron
charge distributions:

\begin{equation}\label{ep005}
        \frac{\partial \,\sigma}{\partial \,t} + 
        \nabla \,\sigma { \vec u} = 0
\end{equation}

the vector field will be equivalent to:

\begin{equation}\label{ep006}
        \nabla ( { \vec j} - \sigma { \vec u}) = 0 
        \quad \Rightarrow \quad
        { \vec j} = \sigma { \vec u}
\end{equation}

Since momentum $ { \vec p} $ can be expressed in terms of $ { \vec j} $,
the wave equation also applies:

\begin{equation}\label{ep007}
        { \vec p} = \rho { \vec u} = 
        \frac{m}{e} \sigma { \vec u} \quad
        \nabla^2 \,{ \vec j} - \frac{1}{u^2} 
        \frac{\partial^2 { \vec j}}{\partial \,t} = 0
\end{equation}

If static problems are considered, $\vec j$ will be independent of time,
and using the definition of magnetic fields as well as 
(\ref{ep002}), the
statement can be rewritten in the following form:

\begin{equation}\label{ep008}
        \nabla \times { \vec B} = 
        \frac{m}{2 e \sigma} \nabla (\nabla { \vec j})
\end{equation}

It is equivalent to {\em Ampere's law}, if electrodynamic current
density is defined by:

\begin{equation}\label{ep009}
        { \vec J} ({ \vec r}, t) 
        \equiv \frac{m c}{8 \pi e \sigma} \nabla ( \nabla
        \sigma\, {\vec u})
\end{equation}

In linear approximation, defining specific resistance 
$ \Omega (\vec r) $ by:

\begin{equation}\label{ep010}
        {\vec E} = \Omega {\vec J} \qquad {\vec E} 
        = - \nabla \phi_{ed}
\end{equation}

and using the averages of density as well as neglecting boundary conditions,
electromagnetic potentials inside a conductor will be due to sources of the
electron wave vector $\vec k$:

\begin{equation}\label{ep011}
        \phi_{ed} =  \frac{m c}{8 \pi e} \nabla { \vec u} =
        \frac{\Omega c}{8 \pi |e|} \nabla \hbar {\vec k}
\end{equation}

Which means, otherwise, that the driving force of electrostatic current 
is the gradient of electron velocity.

\subsection{The system of physical units}\label{ed_units}

A possible way to show, how the system of physical units is 
affected by this, basically field theoretical approach to particle
motion and interaction, is the following: by modeling a framework 
of particle motion consistent with classical electrodynamics, and
comparing the energy expressions derived with energy of electron
waves of hydrogen atoms, the unit of current and the electromagnetic
coupling constants will be defined in terms of motion, they become
{\em mechanical} units. 
If we consider the complementary fields of an electron or photon,
we may describe their features by a vector field 
$ \vec A  (\vec r, t) $, which comprises, from an electrodynamic 
point of view, the dynamical situation:

\begin{eqnarray}
        \psi ({\vec r}, t) \, := \, exp \,i({\vec k} {\vec r} - \omega t)
        \qquad {\vec n} \, := \, \frac{\vec E}{|\vec E|} 
        \nonumber\\
        {\vec A} ({\vec r}, t) := \beta_{f} \frac{\vec n \psi}{e}
        \label{ep012}
\end{eqnarray}

Then electromagnetic fields are, similar to the formulations in 
classical electrodynamics, given by:

\begin{equation}\label{ep013}
        {\vec E} = - \alpha_{E}
        \frac{\partial {\vec A}}{\partial  t} \qquad
        {\vec B} = + \alpha_{B} 
        \nabla \times {\vec A}
\end{equation}

The constants $ \alpha $  have to be added for reasons of
consistency, since the traditional coupling of electromagnetic fields and
dynamical variables cannot be taken for granted.
According to the definitions of $ {\vec E} $ and $ {\vec B} $ in the 
previous section, the $ \alpha $ will only be
equal to unity, if $ [{\vec A}] $ is equal to 
$ \left[\rho {\vec u} / \bar \sigma \right] $.

Based on these definitions and the wave equation, Maxwell's
equations were accounted. If, on the other hand,
$ [{\vec A}] $ is, like in classical electrodynamics, equal to energy
density (see Eq. \ref{ed020}), then it
equals $ {\vec p} c $, which has the dimensions of energy density
but the direction of momentum.
For reasons of consistency of the framework with classical
electrodynamics the constants have to be set to unity and
the vector field shall be equal to momentum per unit charge.
$ \beta_{f} $  then must be dimensionally 
equal to mechanical momentum.

\begin{displaymath}
        \alpha_{E} = \alpha_{B} = 1 \qquad \Rightarrow \qquad
        [\beta_{f}] = [p] = \left[\frac{kg m}{s}\right]
\end{displaymath}

Energy of the electromagnetic complement in terms of
electromagnetic fields has to be modified accordingly. 
Considering, that the time derivative of $ {\vec A} $ equals:

\begin{displaymath}
        \frac{\partial {\vec A} ({\vec r}, t)}{\partial t} =
        - i \omega {\vec A} ({\vec r}, t) \qquad
        {\vec E}^2 ({\vec r}, t) = - \omega^2 {\vec A} ({\vec r}, t)
\end{displaymath}

And equally, that the energy of a particle is proportional to its
frequency:

\begin{displaymath}
        \phi \propto \omega \qquad \frac{\omega^2}{u^2} \propto \omega  
        \quad \Rightarrow \quad  \frac{1}{u^2}\,{\vec E}^2 \propto \phi
\end{displaymath}

Energy density of electromagnetic complements, based on this definition
of $ {\vec A} $ can be written:

\begin{equation} \label{ep014}
        \phi_{e} = \frac{1}{2}\, \left( \frac{1}{u^2}\,{\vec E}^2
        + {\vec B}^2 \right) 
\end{equation}

The correct relation between frequency and energy is then accounted
for, while the constant $ \beta_{f} $ may be estimated from hydrogen
ground states. 
Using the results of material wave theory for electron motion within a
hydrogen atom ~\cite{HOF96} and considering, that the amplitude of 
electromagnetic as well as kinetic energy must be double its 
kinetic value, $ \beta_{f} $ is given by:

\begin{eqnarray}\label{ep015}
        \left( k_{0} A_{0}\right)^2 &=&
        2 M_{e} (\nu_{0} R_{0})^2 \qquad
        k_{0}^2 = 
        \left(\frac{M_{e} (\nu_{0} R_{0})}{\hbar}\right)^2
        \nonumber\\
        A_{0}^2 &=& \frac{2 \hbar^2}{M_{e}} = \frac{\beta_{f}^2}{e^2} 
        \\
        \beta_{f} &=& e \hbar \sqrt{\frac{2}{M_{e}} } =
         2.50 \times 10^{- 38} 
        \left[A m^2 kg^{1/2}\right] \\
        \hbar &=& \beta_{f} \sqrt{\frac{M_{e} }{2 e^2}}
\end{eqnarray}

The unit Ampere in this case is no longer free for definition, but must
be expressed in terms of dynamical units m, kg, s.

\begin{equation}\label{ep016}
        [A] = \left[\frac{kg^{1/2}}{m s}\right]
\end{equation}

Especially interesting is the meaning of coupling constants in
this context.

\begin{eqnarray}
        \left[{\bf \epsilon}^{- 1}\right] = \left[\frac{m^2}{s^2} m^3\right] =
        \left[u^2\, V\right] = \left[M u^2\, \frac{V}{M}\right]
        \nonumber\\
        \left[\mu\right] = \left[V\right] \qquad 
        \left[\alpha_{f}\right] = \left[{\bf \epsilon}^{- 1}\right] = 
        \left[M u^2\, \frac{V}{M}\right]\label{ep017}
\end{eqnarray}

where $ \alpha_{f} $ is Sommerfeld's fine structure constant. The 
units signify, that the coupling constants actually describe
the capacity of electron matter to store dynamical energy per unit
mass: they give rise to the question, treated in the following sections,
how quantum theory allows for energy or charge quantization.
The field corresponding to electromagnetic fields $ \vec E $ and $ 
\vec B $ and in longitudinal direction will be defined by:

\begin{equation}\label{ep018}
        \vec{\pi} ({\vec r}, t) = - i \hbar \nabla \psi
\end{equation}

It is not equal to intrinsic momentum, since it is harmonic in
longitudinal direction. For the plane wave 
$ \psi = exp \,i({\vec k r} - \omega t) $, the fields will
be:

\begin{eqnarray*}
        {\vec E} &=&  \beta_{f} \,{\vec n}\, 
        \sin ({\vec k \vec r} - \omega t) \\
        {\vec B} &=&  \beta_{f} ({\vec k} \times {\vec n}) 
        \sin ({\vec k \vec r} - \omega t) \\
        \vec{\pi} &=& \hbar {\vec k} 
        \cos ({\vec k \vec r} - \omega t)
\end{eqnarray*}

\subsection{Dynamic interactions}

On a more general basis we proceed from the Lagrangian of electron
photon interaction in the presence of an external field $ \phi $. 
The procedure is pretty standard: defining the Lagrange density of 
a particle in motion, including external potentials and a presumed
photon, a variation with fixed endpoints allows to calculate photon
energy density in terms of the external potential and kinetic potential
of the particle, as well as accelerations of the particle due to the
potential applied. So far the result is the traditional one. But
using a Legendre transform for small velocity variations and 
calculating the Hamiltonian density will show, that the 
interaction Hamiltonian is equal to photon energy, which will
allow, on a very general basis, to refer all electrostatic
interactions to an exchange of photons.
We define the Lagrangian by the total energy of the moving
electron, and include total energy of a presumed photon:

\begin{equation}\label{ep019}
        {\cal  L} \, := \, T - V = 
        \rho_{el}^{0} {{\dot{x}}_{i}^2} + \rho_{ph}^{0} c^2 - 
        \sigma_{0} \phi
\end{equation}

\begin{eqnarray*}
        \rho_{el}^{0} &=& constant \quad
        \dot{x}_{i}^2 = \dot{x}_{i}^2 (\dot{x}_{i})\\
        \rho_{ph}^{0} &=& \rho_{ph}^{0} (\dot{x}_{i})\quad
        \phi = \phi (x_{i})
\end{eqnarray*}

Using the Hamiltonian principle for infinitesimal variations of
the system with fixed endpoints:

\begin{eqnarray}
        &\delta& W = \delta \int\limits_{t_{1}}^{t_{2}} dt 
        \int d^3 x {\cal L} = \nonumber \\
        &=&
        \int\limits_{t_{1}}^{t_{2}} dt \int d^3 x \left\{
        \left( \rho_{el}^{0} 
        \frac{\partial \dot{x}_{i}^2}{\partial \dot{x}_{i}}
        + c^2 \frac{\partial \rho_{ph}^{0}}{\partial \dot{x}_{i}} \right) 
        \delta \dot{x}_{i} - \sigma_{el}^{0}\, \frac{\partial \phi}{\partial x_{i}} \, 
        \delta x_{i} \right\}\nonumber
\end{eqnarray}

taking into account that

\begin{displaymath}
        \int\limits_{t_{1}}^{t_{2}} dt 
        \rho_{el}^{0} \frac{\partial \dot{x}_{i}^2}{\partial \dot{x}_{i}}
        \delta \dot{x}_{i} =
        \rho_{el}^{0}  
        \underbrace { \left[\frac{\partial \dot{x}_{i}^2}{\partial \dot{x}_{i}}
        \delta x_{i} \right]_{t_{1}}^{t_{2}} }_{= \,0} -
        \int\limits_{t_{1}}^{t_{2}} dt \frac{d}{d t} \left(
        \frac{\partial \dot{x}_{i}^2}{\partial \dot{x}_{i}} \right) 
        \delta x_{i} 
\end{displaymath}

the integral may be rewritten to:

\begin{equation}\label{ep020}
        \int\limits_{t_{1}}^{t_{2}} dt \int d^3 x \left\{ 
        \sigma_{el}^{0} \,\frac{\partial \phi}{\partial x_{i}} 
        + \frac{d}{d t} \left( \rho_{el}^{0}
        \frac{\partial \dot{x}_{i}^2}{\partial \dot{x}_{i}}
        + c^2 \frac{\partial \rho_{ph}^{0}}{\partial \dot{x}_{i}}
        \right)\right\} \delta x_{i}  = 0
\end{equation}

Therefore the following statement is valid:

\begin{equation}\label{ep021}
        \rho_{ph}^{0}c^2 = - \int dt \underbrace{
        \sigma_{el}^{0} \dot{x}_{i} \nabla \phi 
        }_{=\, - {\vec j}_{0} {\vec E}\,  = 
        \,-  \frac{\partial V}{\partial t} }
        - \rho_{el}^{0} \dot{x}_{i}^2
        = \,+ \,V - \rho_{el}^{0} \dot{x}_{i}^2
\end{equation} 

The dynamic relation in the absence of a photon will be:

\begin{equation}\label{ep022}
        \rho_{el}^{0} \frac{d \dot{x}_{i}}{d t} = 
        - \frac{\sigma_{el}^{0}}{2}\, \nabla \phi
        = - \bar{\sigma}_{el} \,\nabla \phi
\end{equation}

The Legendre transform of the electron photon system in an
external field yields:

\begin{equation}\label{ep023}
        \frac{\partial L}{\partial \dot{x}_{i}} =  
        \frac{\partial V}{\partial \dot{x}_{i}}
\end{equation}

For small variations of velocity 
the Hamiltonian of the system is given by:

\begin{equation}\label{ep024}
        H = \frac{\partial L}{\partial \dot{x}_{i}} \dot{x}_{i}
        - L = \sigma_{el}^0 \enspace \phi
\end{equation}

The result seems paradoxical in view of kinetic energy of the
moving electron, which does not enter into the Hamiltonian.
Assuming, that an inertial particle is accelerated in an 
external field, its energy density after interaction with
this field would only be altered according to its alteration of
location. The contradiction with the energy principle is
only superficial, though. Since the particle will have been
accelerated, its energy density {\em must} be changed. If
this change does not affect its Hamiltonian, the only
possible conclusion is, that photon energy has equally
been changed, and that the energy acquired by acceleration
has simultaneously been emitted by photon emission.
The initial system was therefore over--determined, and
the simultaneous existence of an external field {\em and}
interaction photons is no physical solution to the 
interaction problem.

The process of electron 
acceleration then has to be interpreted as a process of simultaneous 
photon emission: the acquired kinetic energy is balanced
by photon radiation. 
A different way to describe the same result, 
would be saying that electrostatic interactions are
accomplished by an exchange of photons: the potential of
electrostatic fields then is not so much a function of
location than a history of interactions.
This can be shown by calculating the Hamiltonian of electron 
photon interaction:

\begin{eqnarray}
        H_{0} = \rho_{el}^{0}\, \dot{x}_{i}^2 + \sigma_{el}^{0} \phi
        \qquad
        H = \sigma_{el}^{0}\, \phi \nonumber\\
        H_{w} = H - H_{0} = - \rho_{el}^{0} \,\dot{x}_{i}^2 = \rho_{ph}^{0} \,c^2
        \label{ep025}
\end{eqnarray}

\subsection{Structural fields}

Since material wave theory is based on the intrinsic features of
particles, it has to account for a problem, usually treated rather
sloppily in quantum theory. We refer to the problem of particle
scattering.

If Einstein's notion of energy as an intrinsic quality of 
motion is taken seriously - and the present paper seeks to 
accomplish the same for momentum: i.e. to establish momentum 
$\vec p$ as an intrinsic quality of particle motion - then a 
thorough analysis of scattering processes raises problems 
neither quantum theory nor electrodynamics are qualified to solve. 
 
We assume two electrons of integral momenta 
$ {\vec P}_{0}^{1}  $ and $ {\vec P}_{0}^{2}  $   
to interact; during this process their path will be changed
due to electrostatic interactions 

\begin{equation} \label{ep026} 
        \nabla {\vec E}^{i} = \frac{\sigma^{i}}{\epsilon_{0}} \qquad 
        \frac{\partial {\vec P}^{j}}{\partial t} = Q^{j} {\vec E}^{i} 
\end{equation} 

as well as their velocity. The decisive question, in this 
context, is the change of internal structures given by: 

\begin{equation} \label{ep027} 
        \sigma^{i} \left( {\vec r},t \right) = 
        2 \bar{\sigma} {\sin ^2 \left( \frac{{\vec P}^{i}}{\hbar} {\vec r} 
        - \omega^{i} t \right)}  
\end{equation} 

where $ \omega^{i}  $ as well as $ {\vec P}^{i}  $  
are functions of time. 
 
Electric and magnetic intensity $ {\vec E}  $  and  $ {\vec B}  $, 
and their relations were 
shown to be consequences of the internal structure of any 
single electron, electrodynamics in its traditional form therefore 
provides no answer to the question, by what means the 
alteration of internal structure is brought about. 
 
The argumentation of quantum theory in this respect is, strictly 
speaking, illegitimate, because it is not developed beyond 
the simple statement, that a specific momentum causes a 
specific wave packet. By what means this change is 
accomplished, remains unanswered. 
 
From the viewpoint of field theory, the medium of 
transmission can only be a field, whatever its qualities may 
be, and this field, by way of physical interactions, must 
cause the internal restructuring of the second 
particle 

From the definition of electric fields Eq. \ref{ed004} and the 
equation of continuity for electron momentum and density, it may
be concluded that (in a one--dimensional simplification):

\begin{eqnarray}\label{ep028}
\bar \sigma \vec E =: - \nabla \phi_{ext} =
- \nabla \phi_{int} + \frac{\partial \vec p}{\partial t} 
\end{eqnarray}

The time--derivative, including the equation of continuity,
then yields:

\begin{eqnarray}\label{ep029}
\nabla \frac{\partial}{\partial t} \phi_{ext} =
u^2 \triangle \vec p - \frac{\partial^2 \vec p}{\partial t}
\end{eqnarray}

Identifying the external potential as the potential of
a photon:

\begin{eqnarray} \label{ep030}
\phi_{ext} = \rho_{ph} c^2 \qquad
\frac{\partial \phi_{ext}}{\partial t} = - c^2 \nabla \vec p_{ph}                     
\end{eqnarray}

the source term of the electron wave equation for any of
the two particles can be determined:

\begin{eqnarray}\label{ep031}
u^2 \triangle \vec p - \frac{\partial^2 \vec p}{\partial t} =
- c^2 \triangle \vec p_{ph}
\end{eqnarray}

As quantization arises from the interaction of single points 
within the considered region (see the following sections), 
conservation of electron momentum for the two particles and,
consequently, for the two photons, must hold in a very general
sense for an arbitrary location within the two photons. And
using, again, the equation of continuity for intrinsic momenta
and potentials of the photons, the sum of the two potentials
must comply with:

\begin{equation} \label{ep032} 
        \frac{\partial}{\partial t} 
        \left( \phi_{ph}^{1} + \phi_{ph}^{2} \right) = 0 
\end{equation} 

If the constant of integration equals zero, the total field 
of the two photons will vanish for every external observer: 

\begin{equation} \label{ep033} 
        \phi_{ph}^{1} + \phi_{ph}^{2} = \mbox{constant} = 0 
\end{equation} 

A specific feature of a component $ \phi_{ph}^{i}  $ 
of the overall field will be, that it 
does not depend on the distance from its source 
$ \nabla {\vec p}^{i}  $: 

\begin{equation} \label{ep034} 
        \phi_{ph}^{i} \not= f(|{\vec r} - {\vec r}^{i}|) g^{i} 
\end{equation} 

It is a consequence of  (\ref{ep031}), 
because if  intensity would in  
some way depend on the distance  $ r $, 
linear momentum could not remain constant for the 
system of two particles. 

If the external and static field $ - \nabla \phi $ is the field of an
interacting particle, then emission and absorption will be balanced.
The traditional Hamiltonian, in this case, correctly describes dynamical 
variables of interaction, although only due to the - hidden - process
of photon exchange.
It is evident that neither emission, nor absorption of photons at this stage
does show discrete energy levels: the alteration of velocity can be
chosen arbitrarily small.
The interesting question seems to be, how quantization fits  
into this picture of continuous processes. 
 
\subsection{Electrodynamic fields and energy quantization} 

To understand this paradoxon, we consider the  
transfer of energy due to the infinitesimal interaction 
process. The differential of energy is given by: 

\begin{equation} 
        d\,E = A\cdot\,dt\,\rho_{ph}^{0}\,c^2 
\end{equation} 

A denotes the cross section of interaction. Setting $ dt $ 
equal to one period $ \tau $, and considering, that the  
energy transfer during this interval will be $ \alpha\,E $, 
we get for the transfer rate: 

\begin{equation} 
        \alpha\,\frac{d\,E}{d t} = 
        \frac{A\,\lambda\,\rho_{ph}^{0}\,c^2}   
        {d t} = \alpha\cdot\frac{V_{ph}\,\rho_{ph}^{0}\,c^2}{\tau} = 
        \alpha\, \frac{h\,\nu}{\tau} 
\end{equation} 

Since total energy density of the particle as well as the 
photon remains constant, the statement is generally valid. 
And therefore the transfer rate in interaction processes will 
be: 

\begin{equation} 
        \frac{d\,E}{d t} = \frac{d}{d t}\,M\,{\vec   u}^2 = 
        \hbar\cdot\frac{\omega}{\tau} = h\cdot\nu^2 
\end{equation} 

The result was already quoted (see section \ref{sec_ed}). And it
was equally emphasized, that the term energy quantum is, from
the point of view taken in material wave theory, not quite
appropriate. The total value of energy transferred depends, in
this context, on the region subject to interaction processes, and
equally, as will be seen further on, on the duration of the
emission process. The view taken in quantum theory is therefore
only a good approximation: a thorougher concept, of which 
material wave theory in its present form is only the outline,
will have to account for {\em every} possible variation in the
interaction processes.

But even on this, limited, basis of understanding,
Planck's constant, commonly considered the fundamental value 
of energy  quantization  is the fundamental constant not of 
energy values, but of transfer rates in dynamic processes. 
Constant transfer rates furthermore have the consequence, that 
volume and mass values of photons or electrons become 
irrelevant: the basic relations for energy and dispersion 
remain valid regardless of actual quantities. 
Quantization then is, in short, a {\em result of energy transfer} and 
its characteristics. 

\begin{itemize}
\item
If intrinsic features of particles and their interactions
are considered, quantization has to be interpreted as a
result of energy transfer processes and their specific properties.
\end{itemize}

\subsection{Electrostatic fields and charge quantization}

The photon interaction process leading to the acceleration of particles in
an electrostatic field is accounted for by the field
vector $ {\vec A}_{L} ({\vec r}) $, of longitudinal orientation
and including the effects of different field polarizations: 

\begin{equation}
        \int\limits_{S} d \Omega\, n_{i} n_{j} = 
        \frac{4 \pi}{3} \delta_{i j} \quad 
        {\vec A}_{L} ({\vec r}) \equiv 
        \frac{4 \pi}{3} \frac{\beta_{f}}{e}  
        \psi ({\vec r}\, , {\vec k} \rightarrow 0) \,{\vec e}^{r}
\end{equation}

Elementary charge e shall be the flow through a surface 
of the field $ {\vec A}_{L} ({\vec r}) $ 
in unit time $ t_{1} - t_{0} = 1 $ :

\begin{equation} \label{ch001}
        e \equiv  \frac{d}{d \Omega}\, \int\limits_{t_{0}}^{t_{1}}d t\,
        \nabla\, {\vec A}_{L} ({\vec r})
\end{equation}

Integrating over a spherical surface with r = 1 and evaluating the
time integral yields:

\begin{eqnarray}
        \int\limits_{V} d^3 r' e &=& \frac{4 \pi}{3} e = \frac{d}{d \Omega}
        \int\limits_{t_{0}}^{t_{1}} d t \int\limits_{V} d^3 r' 
        \nabla {\vec A}_{L} \\
        &=& \int\limits_{t_{0}}^{t_{1}} d t 
        \int\limits_{S} d^2 {\vec f} 
        \frac{d}{d \Omega} \,{\vec A}_{L}
        = (t_{1} - t_{0}) \,{\vec A}_{L} \nonumber 
\end{eqnarray}

The result is consistent with the source relation of electrostatic fields,
since (in SI units):

\begin{eqnarray*}
        \nabla {\vec D} &=& \nabla\, \epsilon_{0} {\vec E} 
        = \delta ({\vec r}) e
        \nonumber \\
        \frac{d}{d t} \oint\limits_{S} d^2 {\vec f}\, {\vec D}
        &=& \underbrace{
        \oint\limits_{S} d^2 {\vec f}\,\frac{d}{d \Omega} {\vec A}_{L} 
        \delta ({\vec r'}) }_{=\, 0}
        - \frac{d}{d \Omega} 
        \int d^3 r'\, {\vec A}_{L} \nabla \delta ({\vec r'})
\end{eqnarray*}

\begin{equation}
        {\vec D} = - \frac{ \frac{d}{d \Omega}
        \int\limits_{t_{0}}^{t_{1}}d t \,\nabla\, {\vec A}_{L} 
        ({\vec r}) }{4 \pi \,r^2} {\vec e}^{r} 
\end{equation}

And the flow in unit time per unit volume is then:

\begin{eqnarray}
        (t_{1} - t_{0}) \,{\vec A}_{L} = 
        \,{\vec A}_{L} = \frac{\beta_{f}}{e} \frac{4 \pi}{3}
        = \frac{4 \pi}{3} e \nonumber\\
        \beta_{f} = e^2 \qquad e = \sqrt{\beta_{f}}
\end{eqnarray}

The calculation yields the numerically correct result. 
It is also dimensionally consistent, if it is considered, that
it refers to unit volume and in unit time.
The elementary quantum of charge is therefore not exactly a
quantity of some physical variable, but, consistent with
previous results, a measure for dynamical energy transfer rates
per unit time.
Charge quantization is the consequence of {\em photon
interaction processes and their specific energy transfer rates}. 

\begin{itemize}
\item
The elementary quantum of electron charge derives from 
energy transfer rates in electrostatic interactions, which
are an expression of the interaction process accomplished
by photons.
\end{itemize}

\subsection{Planck's constant and quantum theory}

Using the result for $ \beta_{f} $ in terms of e,
Planck's constant equally is not a fundamental constant but
can be referred to dynamic interaction rates of particles
and fields (dimensions again corrected):

\begin{eqnarray}
        \hbar &=& \beta_{f} \sqrt{\frac{M_{e}}{2 e^2} } = 
        e \sqrt{\frac{M_{e}}{2}} \\
        \hbar &=& 1.081 \times 10^{-34} Js \nonumber
\end{eqnarray}

Comparing with the values for $ \hbar $ based on CODATA recommendation 
for a consistent system of fundamental constants \cite{UIP78}, the
result deviates by about 2.5 \%:

\begin{equation}
        \hbar_{UIP} = 1.054588 \times 10^{-34} Js\qquad
        \triangle \hbar \approx 2.5 \%
\end{equation}

But considering, that the result was obtained by a completely different
approach and, furthermore, that it is the first {\em direct} relation
between the three fundamental constants $e, M_{e}, \hbar$ without any
additional parameter, the numerical value seems in acceptable accordance
with current data.

If the relation holds, then the consequences for the framework of
quantum theory are significant. Its essential conception, the interpretation
of discrete frequency, mass and charge values as discrete quantities
determining physical processes can, from this viewpoint, no longer be
sustained. All processes analyzed so far, were found to be continuous in
nature, and the appearance of discrete interaction quanta had to be
referred to constant transfer rates applying to every single point
of the underlying fields. 

\begin{itemize}
\item
On the basis of electron--photon interactions Planck's constant
in material wave theory only applies to interactions and transfer
processes. A justification of discrete frequencies, mass or charge
values from the viewpoint of allowed {\em states} cannot be given
in material wave theory. The result indicates, that the formalism
of quantum theory lacks justification if it refers discrete 
energy levels to isolated particles.
\end{itemize}

The result provides the logical reason for the deeply rooted
epistemological problem of quantum theory expressed by
Heisenberg (see ref. \cite{REN60}): If quantization is only
appropriate for interactions, i.e. measurement processes, then
the results of quantum theory can only hold for actual 
measurement processes. But since the formalism of quantum 
theory is based on single {\em eigenstates}, meaning states
of isolated particles, this logical structure is not
accounted for by the mathematical foundations of quantum
theory. While, therefore, the mathematical formalism suggests
a validity beyond any actual measurement, it can only be
applied to specific measurements. What it amounts to, in short,
is a logical inconsistency in the fundamental statements of
quantum theory.

\subsection{The problem of charge}

The assumption of invariant charge, i.e. the conception, that charge 
remains invariant regardless of the interaction process or the 
surrounding, is not compatible with the results so far. If charge
can only be determined by its photon interactions with other charge
-- a result of the previous sections -- and if the term charge
refers to the {\em electromagnetic} qualities of this interaction
-- as described by the change of kinetic variables of charged 
particles -- then the {\em charge} of a particle must be referred
to the dynamic qualities of interaction photons. 

As the dynamic qualities of a photon (not accounting for
transversal properties which will be treated in the next section)
are completely determined by (1) its frequency $\nu$ and (2) its
momentum $ {\vec p} = \hbar {\vec k} $, the possible interactions
of two particles of mass $m$ and charge $\pm$ e must be 
accomplished by a photon with linear momentum ${\vec p}$ 
described by (propagation assumed parallel to ${\vec e}_{x})$:

\begin{equation}
        {\vec p} = {\vec p}_{0} \sin^2 (\pm k x - \omega t)
\end{equation}

The energy of the photon in every case equals
$ \hbar \omega $, both forms comply with the wave equation
for linear intrinsic momentum and yield identical transversal
fields, the difference, though, is
that the form $ {\vec p}_{0} \sin^2 (- k x - \omega t)$ 
cannot be interpreted as a particle, because its momentum
variable has the opposite direction of propagation. 

\begin{itemize}
\item
The interpretation of electrostatic interactions
as an exchange of photons, proper to material wave theory,
has the consequence, that (1) the properties of single
charges can only be referred to interactions with the
environment, (2) photons can only be interpreted as particles
if two identical charges interact, and (3) electrostatic
potentials depend on the possibility of photon exchanges.
\end{itemize}

While for two separate charges the third consequence is of
no effect, it cannot be assumed a priori, that aggregations
of charge typical for nuclear environments will provide a
mode of photon interactions. But if atomic nuclei do not
provide a mode of photon interaction between individual nuclear
''particles'', then electrostatic considerations do not
apply to the problem of nuclear stability.

\subsection{Polarization}

So far the vector characteristics of electromagnetic fields
have not been accounted for.
The problem of electron photon interaction in this case
requires a closer scrutiny of the energy transfer process itself.
Consistent with classical electrodynamics it may be postulated,
that electromagnetic fields may be superimposed. In this case
the vector field $ \vec{A} $ of an electron and interacting photon
is given by:

\begin{equation}
{\vec A} = {\vec A}_{el} + {\vec A}_{ph}
\end{equation}

And consistent with the framework developed it may be stated, that
total energy density at any location of the electron must be equal
to:

\begin{equation}
\rho_{el}^{0}\, u^2 = \frac{1}{2} \,
\left( \frac{1}{u^2} {\vec E}_{0}^2 + {\vec B}_{0}^2 \right)
\end{equation}

Using (\ref{ep014}) and (\ref{ep015}) the energy relation yields:

\begin{eqnarray}
\hbar^2 {\vec k}^2 =  
\hbar^2 {\vec k}_{el}^2 + \hbar^2 {\vec k}_{ph}^2
+ 2 \hbar^2 {\vec k}_{el} {\vec k}_{ph} \nonumber\\
{\vec k}^2 = 
{\vec k}_{el}^2 + {\vec k}_{ph}^2
+ 2 |{\vec k}_{el}| |{\vec k}_{ph}| \cos \vartheta \\
\vartheta = \vartheta (\vec E_{el}, E_{ph}) \nonumber
\end{eqnarray} 

The energy is altered depending on the angle between 
electromagnetic field vectors of electron and interacting
photon. The equation has the same structure as the
energy relation in classical mechanics, it does not refer, though,
to any longitudinal characteristic. Even if longitudinal
vectors are parallel, the difference in transversal fields
must account for energy deviations.
Taking two identical $k$--vectors and assuming, that every angle
of deviation is equally probable, then the $k$--vector after
interaction will be in average:

\begin{equation}
{\vec k}^2 = 2 {\vec k}_{i}^2 \left(1 \pm \frac{2}{\pi} \right)
\end{equation}

And if $ 2 {\vec k}_{i}^2 $ is considered as the undisturbed 
interaction energy, then the average deviations are given by:

\begin{equation}
\triangle W = \pm \frac{2}{\pi}\, W_{0} = \pm \frac{2}{\pi}\,
\hbar \omega
\end{equation}

To estimate the importance of polarizations consider two electrons
with arbitrary initial velocities and different orientation of 
intrinsic electromagnetic fields to interact. Then the 
electromagnetic fields of the photons
of interaction will show successively shifted angles and the
kinetic characteristics of particle motion will be affected in
a way not described by electrodynamics. It is this process, which
must ultimately lead to a shift of scattering cross--sections
already predicted and confirmed by quantum theory. 

As a mathematical framework to calculate scattering processes
on a dynamical basis -- complicated by the fact that the standard
scattering approximations are bound to fail, since no potential
can be defined for the angular dependency of photon interaction 
-- has not yet been developed, the exact changes compared to
quantum theory cannot currently be exhibited. 

\subsection{Particle ''spin''}\label{ep_spin}

David Bohm devised a gedankenexperiment to differ between the
theory of hidden variables and quantum theory \cite{BOH51}, which
was based on the assumption, that the direction of spin is a
true but hidden variable. As Bell formulated in his inequalities,
the consequence of such an assumption leads to consequences, which should
be, in principle, measurable \cite{BEL66}. Measurements of these
consequences by Aspect seemed to confirm, that a local 
theory of hidden variables is contradictory \cite{ASP83}. 
Any theory, thus the result, which assumes hidden variables of
this type, is contradicted by measurements. 

This conclusion is not undisputed, as Thompson recently showed by the
''chaotic ball'' model of this type of measurements \cite{THO96}.
The usual interpretation, the author concluded, which assumes
quantum theory to be superior to local hidden variable theories,
contains a biased interpretation of experimental data.

The theoretical framework of material waves, which can be seen as 
an extension of electrodynamics to intrinsic particle properties,
is basically a local theory. Particle spin is a local variable,
and the theory is therefore a realistic and local
theory of ''hidden variables'',
although, and this seems to be an important difference to Bohm's
concept, classical electrodynamics {\em and} quantum theory were
shown to result from specific limitations.
On these grounds the conflict of a
measurement presumably excluding hidden variables in a local
theory and a concept which explicitly goes beyond the limits of
formalization of the quantum theory, on which evaluation of these
measurements is based (like the current framework), can only be
clarified by exclusion. If these measurements are also in the
current framework valid {\it within} the limits of quantum theory, then
the current framework must be ruled out, since it contradicts
experimental evidence of a highly consistent and successful theory.
But if, on the other hand, these measurements do not yield valid
results within the framework of quantum theory, then this experiment
is not conclusive, because its interpretation presupposes theoretical
formulations which have been violated. In this case an alternative
interpretation of the measurements is possible, which is based
on a different theory. 
Although it cannot currently be determined {\em what} was
actually measured in Aspect's experiments, it can be excluded
that the measured result was obtained within the limits of
quantum theory. And since Bell's inequalities, which provided
the theoretical background of interpretation, are based on the
axioms of quantum theory, the interpretation of these results 
is questionable. We will return to these results in later publications.

In the present framework spin is local and parallel to the 
polarization of the intrinsic magnetic field $\vec B$ which can be 
shown as follows:

The energy relations in classical electrodynamics yield, for
a mass of magnetic moment $\vec \mu$ in a magnetic field $\vec B$,
the following energy:

\begin{eqnarray}\label{ep051}
W = - \vec \mu \cdot \vec B
\end{eqnarray}

The following deduction is based on four assumptions:
(1) The energy of an electron or photon is equal to the energy
of the particle in quantum theory, (2) the magnetic field
is the intrinsic (photon) or external (electron) magnetic field,
(3) the frequency $\omega$ is also a frequency of rotation, and
(4) the magnetic moment is described by the relations in quantum 
theory. With (1) the energy of photons (total energy) and electrons 
(only kinetic energy) will be:

\begin{eqnarray}\label{ep052}
W_{ph} = \hbar \omega \qquad
W_{el} = \frac{1}{2} \hbar \omega  
\end{eqnarray} 

With (2) the magnetic fields of electrons and photons will be
(see Eq. \ref{ep001}):

\begin{eqnarray}\label{ep053}
\vec B_{el} &=& - \frac{1}{2 \bar \sigma} \nabla \times \vec p
= \frac{\rho}{\bar \sigma} \vec \omega \nonumber \\
\vec B_{ph} &=& - \frac{1}{\bar \sigma} \nabla \times \vec p
= \frac{2 \rho}{\bar \sigma} \vec \omega
\end{eqnarray}

And with (4) the magnetic moment will be ($g_{s}$ denotes the
gyromagnetic ratio):

\begin{eqnarray}\label{ep054}
\vec \mu = g_{s} \frac{e}{2 m c} \vec s 
\end{eqnarray}

\subsubsection{Bosons}

Calculating the energy of interaction and considering, that the
magnetic field in the current framework is $c$ times the magnetic
field in classical electrodynamics (see Eq. \ref{ed009}), we get
for the photon:

\begin{eqnarray}\label{ep055}
W_{ph} = \hbar \omega = g_{ph} \frac{e}{2 m}
\frac{2 \rho}{\bar \sigma} \vec \omega \vec s_{ph}
\end{eqnarray}

With $ \rho/\bar \sigma = m/e $, which follows from the 
definition of magnetic fields (see section \ref{ed_units}), the
relation can only hold for constant $g_{ph}$ if:

\begin{eqnarray}\label{ep056} 
\vec s_{ph} \parallel \vec \omega \quad 
\vec s_{ph} \cdot \vec \omega = s_{ph} \, \omega \nonumber \\
\Rightarrow \qquad \hbar = g_{ph} s_{ph}
\end{eqnarray} 

The energy relation of electrodynamics in this case is only 
consistent with the energy of photons of they possess intrinsic
spin $\hbar$, a gyromagnetic ratio of 1, and if the direction
of spin--polarization is equal to the direction of intrinsic
magnetic fields. A consistent framework therefore requires 
(1) a localized spin, and (2) a spin $\hbar$ of bosons.

\subsubsection{Fermions}

The same calculation for the properties of electron spin yields
the following result:

\begin{eqnarray}\label{ep057} 
\vec s_{el} \parallel \vec \omega \quad 
\vec s_{el} \cdot \vec \omega = s_{el} \, \omega \nonumber \\
\Rightarrow \qquad \hbar = g_{el} s_{el}
\end{eqnarray} 

In principle the spin variable and the gyromagnetic ratio could
be chosen like the values for photons. From the condition itself,
the difference between bosons and fermions is not directly
accessible. The difference can be accounted for, if the atomic
model in quantum theory is considered: the theory of hydrogen 
fine structures by Uhlenbeck and Goudsmit \cite {UHL26} requires 
two energy levels symmetric to the undisturbed one, thus a
general multiplicity of 2. The only solution, with this
initial assumption, is a spin variable of $\hbar/2$ and a 
gyromagnetic ratio of 2.
The direction of spin polarization is also for fermions equal to
the direction of the magnetic field vector.

Generally the concept of spin seems questionable, because the
frequency $\omega$ is not at all to be confused with rotation
frequencies of a particle. Consider, for example, a photon in
uniform motion $c$ along a flat space--time geometry. Its 
intrinsic magnetic field does not determine any state of
rotation, nor does the energy of a photon result from the
interaction of a photon magnetic moment with a magnetic field. 
From the given deduction of spin--properties of particles 
(and the deduction seems highly appropriate, considering the
effect of spin operators in the framework of quantum theory)
it seems, that the concept itself must be theoretically
deficient.

\subsection{Bell's inequalities}

\begin{figure}
\epsfxsize=1.0\hsize
\epsfbox{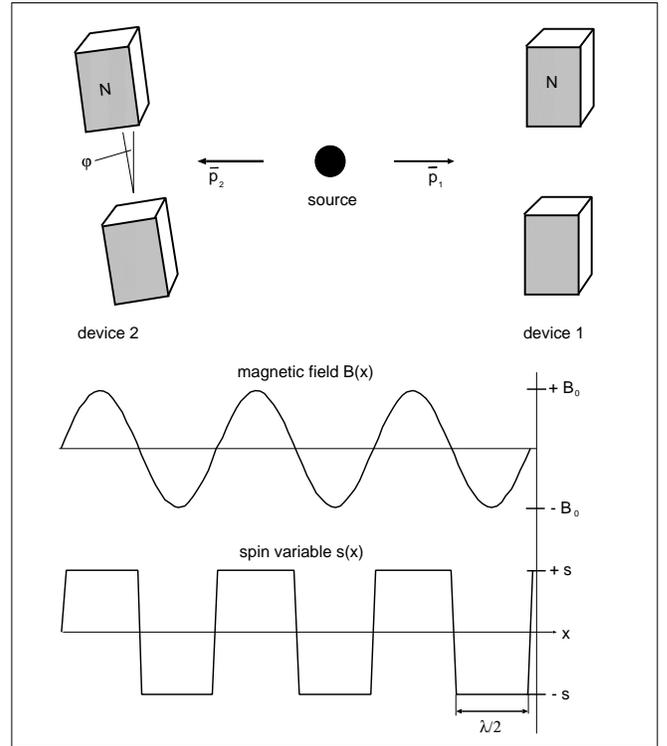} 
\caption{EPR measurement of spin variable. Two Stern--Gerlach
devices (1,2) in different orientations used to determine the 
spin eigenvalues of the particles $\vec p_{1}$ and $\vec p_{2}$ 
from the source (top). The magnetic fields and spin variables 
depend on the position of measurement. Note that the spin 
variables are only constant in an interval shorter than 
$\lambda/2$ (bottom)}
\label{fig001}
\end{figure}

Let us consider, on this basis, the result of any realistic
measurement of photon ''spin'' in an EPR--like situation
\cite{THO96}. The first problem, we are confronted with, is
the exact definition of the spin value. From the given deduction
it seems evident, that the local direction of the spin vector is
either parallel or antiparallel to the intrinsic magnetic field.
Provided, we adopt the view of quantum theory for spin
conservation of two photons, then spin--up would mean, that
the vector is antiparallel to $\vec B$, spin--down means,
it must be parallel. In this case we have two photons with
an arbitrary angle $\vartheta$ of polarization in the state
+ 1 for photon one and -1 for photon two (see Fig. \ref{fig001}).  

The second problem originates from the periodic features of
the intrinsic magnetic field. Since the magnetic field vector
is periodic with 
$\vec B = \vec B_{0} \cos(\vec k \vec r - \omega t)$, the
spin variable {\em cannot} remain constant, but must
oscillate from + 1 to - 1. This oscillation requires, that
we define the exact moment of measurement to obtain any
valid result on the state of either particle. If the
photon path remains constant in measurements, this requirement
is fulfilled, but if, like in Aspect's measurements \cite{ASP83},
the measuring apparatus has to be rotated to account for 
different angles of observation, the effect on measurements
seems not easy to estimate. This problem therefore leads to
a problem of experimental methods, which is hidden in the
usual argumentation of quantum theory. 

Provided, the problem can be accounted for, our next problem
is more substantial. Since the spin variable oscillates, the
measurement can only yield a definite result, if the variable
is measured in the time--interval $t \le \tau/2$, which requires,
that the local resolution of the measuring apparatus is
below $ \triangle x \le \lambda/2$. Now as demonstrated in 
the deduction of the uncertainty relations, this interval is
equal to the lowest limit of local resolution in the framework
of quantum theory. Thus we get the result, that a valid 
measurement of the spin--variable must have a local 
precision which is explicitly beyond the fundamental
uncertainty in quantum theory. If, therefore, the measurements
yield any correlation related to physical processes, then the
fundamental axioms of quantum theory must be violated.

This result seems to turn the controversy about the EPR
paradox into a new direction. It can be established, that the
mathematical framework of quantum theory, describing the state
of either particle by a superposition of the two possible 
states $\pm$ 1 is correct. But it can equally be established,
that a significant correlation in spin--measurements is not,
as Einstein suspected, a proof of hidden variables or a 
contradiction in terms of special relativity, but a proof
of measurements below the fundamental level of uncertainty.
If this sort of measurement is possible, then the uncertainty 
relations are violated.

\begin{itemize}
\item
If EPR measurements are possible and yield a significant
correlation for a pair of spin particles, then the uncertainty
relations are violated.
\end{itemize} 

\subsection{Compton scattering}

In section \ref{sec_ed} it was established, that electromagnetic
radiation consists of two distinct intrinsic variables: (1) a
longitudinal momentum, and (2) a transversal electromagnetic
field, both variables depending on the frequency of radiation. In this
section we found, that the change of energy density due to the 
adsorption of a photon result in motion of the absorbing electron,
the energy increase given by:

\begin{eqnarray}\label{101}
        \hbar \omega_{ph} = m_{el} u_{el}^2
\end{eqnarray}

If monochromatic x--rays therefore interact with free electrons 
, i.e. electrons with a velocity small compared to 
$ \sqrt{\frac{\hbar \omega_{ph}}{m_{el}}} $, then the absorbing
electrons should move in longitudinal direction with a velocity
equal to:

\begin{eqnarray}\label{102}
        u_{el}^{0} = \sqrt{\frac{\hbar \omega_{ph}}{m_{el}}} 
\end{eqnarray}

If this electron is decelerated, then it will emit electromagnetic
radiation with a frequency described by:

\begin{eqnarray}\label{103}
        \frac{d \phi}{d t} = \frac{1}{V_{el}}h \nu^2
\end{eqnarray}

and, depending on the duration of the deceleration process, the 
radiation can possess arbitrarily low frequency. This {\em primary}
emission therefore will not exhibit discrete frequency levels.
But if {\em secondary} processes are considered, then the frequency 
change will depend on (1) the relative velocity between the source
of radiation and the moving electron, and (2) the emission and
absorption characteristics. Assuming, that the emission and 
absorption processes are symmetric in terms of duration, then
the frequency of secondary emission will depend on the original
frequency as well as the angle of emission.  The model presented
is therefore a simplification, and it cannot be excluded, that
a precise and deterministic picture of the whole process will not
yield a deviating result. But compared to the standard model in
quantum theory \cite{DEB23}, it is nevertheless an improvement,
because it does not have to recur to mechanical concepts.

If the electron moves in longitudinal direction and with a
velocity equal to $u_{el}^0$, the frequency $\nu_{S}$ of the 
source will be changed due to the electromagnetic Doppler 
effect. The secondary absorption process therefore has to
account for a frequency $\nu_{S}'$:

\begin{eqnarray}\label{ep104}
        \nu_{S}' = \nu_{S}\, \sqrt{\frac{1 - \frac{u_{el}^0}{c}}{1 +
        \frac{u_{el}^0}{c}}} \approx \nu_{S}
        \left(1 - \frac{u_{el}^0}{c}\right)
\end{eqnarray}

An observer moving with the velocity $u_{el}^0$ of the electron
will therefore measure changed frequencies and wavelengths of 
secondary emission. The wavelength $\lambda_{S}'$ will be:

\begin{eqnarray}\label{ep105}
        \lambda_{S}' = \lambda_{S}
        \left(1 + \frac{h}{m c} \frac{1}{\lambda_{S}}\right) =
        \lambda_{S} + \frac{h}{m c}
\end{eqnarray}

The second term, the shift of wavelength due to longitudinal 
motion of the electron relative to the source of radiation, 
is the {\em Compton wavelength} of an electron \cite{COM23}.  
It is the general shift of wavelength for every electromagnetic 
field, and an expression of longitudinal motion. 

\begin{equation} \label{ep106} 
        \triangle \lambda \left( u_{el}^0 \right) = 
        \lambda_{C} = \frac{h}{m c} = \mbox{constant} 
\end{equation} 

Additionally it has to be considered, that the reference
frame of observation is the same as the reference frame of
the original radiation, thus a second frequency 
shift depends on the angle of observation $\vartheta$. 
Measured from the direction of field propagation this
additional shift will be:
 
\begin{equation} \label{ep107} 
        u' = u_{el}^0 \cos{\vartheta} 
\end{equation} 

And the overall shift of wavelength measured is therefore: 

\begin{equation} \label{ep108} 
        \triangle \lambda = \triangle \lambda \left( u_{el}^0 
        \right) \left( 1 - \cos{\vartheta} \right) = 
        \lambda_{C} \left( 1 - \cos{\vartheta} \right) 
\end{equation} 

The result is equivalent to Compton's measurements. Compton
scattering, in our view, is therefore a result of (1) secondary
absorption and emission processes, and of (2) successive
Doppler shifts due to longitudinal motion of the emitting
electrons. That classical electrodynamics does not provide
a theoretical model for Compton scattering, can be understood
as a consequence of neglecting longitudinal and kinetic
features of photons. Since electromagnetic radiation in
electrodynamics is interpreted from transversal properties
alone (which are periodic), a uniform longitudinal motion
due to absorption processes is not within its theoretical limits.

The possibility to account for quantum effects with the modified
and still continuous model of electrons is strongly supporting, 
in our view, the validity of the unification of electrodynamics and 
quantum theory by formalizing the intrinsic features of 
particles. 


\section{Conclusion}

As it turns out, retrospectively, the theory of material waves
removes the two fundamental problems, which made a physical
interpretation (as opposed to the probability--interpretation
\cite{BORN26})
of intrinsic wave properties impossible. (1) The contradiction
between phase velocity and mechanical velocity of a particle
wave is removed by the discovery of intrinsic potentials.
(2) The problem of intrinsic Coulomb interactions is removed,
because electrostatic interactions can be referred to an 
exchange of photons. A particle in uniform motion is therefore
stable.

To exhibit the full range of logical and physical
implications of the theory developed,
we repeat the essential results of this essay on material
waves, displayed throughout the preceding sections of this
paper:

\begin{enumerate}
\item
The interpretation of particles as physical waves
allows the conclusion, that quantum theory cannot be, in a logical
sense, a complete theoretical description of micro physical systems.
\item
{\em Because} quantum theory neglects intrinsic characteristics
of particle motion, the Schr\"odinger equation is generally 
not precise by a value described by Heisenberg's uncertainty
relations.
\item
Electrodynamics and wave mechanics are theoretical concepts
based on the same physical phenomena, the intrinsic features of
moving particles.
\item
The wave equations of intrinsic particle structures are
Lorentz invariant only for an integral evaluation of energy.
The total intrinsic potentials increase for every transformation
into a moving frame of reference.
\item
If intrinsic features of particles and their interactions
are considered, quantization has to be interpreted as a
result of energy transfer processes and their specific properties.
\item
The elementary quantum of electron charge derives from 
energy transfer rates in electrostatic interactions, which
are an expression of the interaction process accomplished
by photons.
\item
On the basis of electron--photon interactions Planck's constant
in material wave theory only applies to interactions and transfer
processes. A justification of discrete frequencies, mass or charge
values from the viewpoint of allowed {\em states} cannot be given. 
The result indicates, that the formalism
of quantum theory lacks justification if it refers discrete 
energy levels to isolated particles.
\item
The interpretation of electrostatic interactions
as an exchange of photons, proper to material wave theory,
has the consequence, that (1) the properties of single
charges can only be referred to interactions with the
environment, (2) photons can only be interpreted as particles
if two identical charges interact, and (3) electrostatic
potentials depend on the possibility of photon exchanges.
\item
If EPR measurements are possible and yield a significant
correlation for a pair of spin particles, then the uncertainty
relations are violated.
\end{enumerate}

\begin{figure}
\epsfxsize=1.0\hsize
\epsfbox{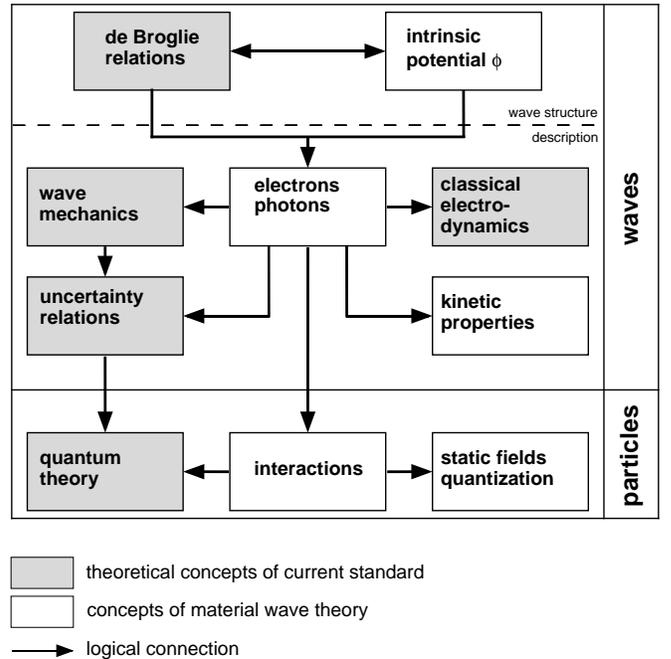} 
\caption{Logical structure of micro physics according to material
wave theory. Classical electrodynamics and wave mechanics describe
intrinsic features of particles, quantization arises from interactions.} 
\label{fig002}
\end{figure}

The theoretical framework thus provides definite answers to the
EPR paradox by rejecting the Copenhagen interpretation of quantum
theory \cite{KOP26}, because (1) quantum theory is not a complete account of
micro physical systems, since the wave function possesses 
physical significance, and (2) the theory is logically
inconsistent, since it is applicable only to interaction processes,
while its mathematical formulation suggests the results to derive
from properties of individual particles. Concerning Renninger's
objections on the quality of photons derived from thought experiments
\cite{REN53}, it provides a model of photons where (1) the energy is
homogeneously distributed over the whole region of electromagnetic 
fields, (2) the kinetic features of photons derive from a longitudinal
component not accounted for in electrodynamics, and (3) the energy
transfer in interactions applies for every single point of a given
region, which yields the quantum effect of interaction.

\section{Discussion}

\subsection{General}

It is difficult to estimate either the impact or the subsequent
changes due to a concept, which completely deviates from the mainstream
of current physics. The aim of establishing and, to a limited extent,
interpreting a physical reality by an alternative framework was to
provide a means, by which experimental results can be seen from
at least two different points of view. The fundamental idea, it
seems, is not very different from the conceptions of Einstein,
Schr\"odinger, Bohm, or de Broglie (see the introduction), 
but the disregard of well established interpretations is 
considerably higher. 

This was not, initially, intended. It only turned out,
that a theoretical concept limited to only minor changes will
neither provide a scientifically satisfying solution -- since
it still is based on some {\em have to believe} -- nor does it
really offer any new insight into micro physical processes.
It was not intended to recover all the 
results of current quantum physics, not least, because it would
surely require many lifetimes to succeed. But it was intended,
and, so I think, accomplished rather successfully, to extend
the theoretical framework of micro physics beyond the
uncertainty relations and the {\em we cannot know} of quantum
theory. Whether the result presented it the best possible way 
out of the dilemma arising from a physics of quantities --
mechanics -- confronted with a world of intensities 
-- electrodynamics -- cannot be currently estimated. 
To estimate the epistemological changes implied by the
theoretical framework, Fig. \ref{fig002} displays the logical
structure of micro physics in material wave theory.

\subsection{Experimental}

To estimate, whether the new theory contradicts any experimental
evidence within its range of formalization, it has to be considered,
that it provides a meta--theory combining the different frameworks
of electrodynamics, quantum theory, and quantum electrodynamics in
a single consistent framework.

From the viewpoint of electrodynamics, it will reproduce any result
the current framework provides, because the basic relations are not
changed, merely interpreted in terms of electron and photon 
propagation. Therefore an experiment, consistent with electrodynamics,
will also be consistent with the new theory. 

In quantum theory, the only statements additional to the original
framework are beyond the level of experimental validity, defined 
by the uncertainty relations. Since quantum theory cannot, in 
principle, contain results beyond that level, every experiment
within the framework of quantum theory must necessarily be
reproduced. That applies to the results of wave mechanics, which
yields the eigenvalues of physical processes, as well as to the
results of operator calculations, if von Neumann's proof of the
equivalence of Schr\"odinger's equation and the commutator 
relations is valid. The only result, which could in this respect
disprove the theory, is the spreading of a wave packet: but as 
this experimental result was the reason, that a {\em physical}
interpretation of wave functions had to be discarded, the theory
also in this respect seems to stand on firm ground.

In quantum field theory, the same rule applies: the new theory only
states, what is, in quantum theory, no part of a measurement
result, therefore a contradiction cannot exist. If, on the other
hand, results in quantum electrodynamics exist, which are not yet
verified by the new theory, then it is rather a problem of 
further development, but not of a contradiction with existing
measurements.

\subsection{Theoretical}

Experimentally, the new theory cannot be disproved by existing
measurements, because its statements at once verify the existing
formulations and extend the framework of theoretical calculations.
The same does not hold, though, for the verification of existing
theoretical schemes of calculation by the new framework. Since the
theory extends far beyond the level of current concepts,
established theoretical results may well be subject to revision. The
theory of material waves requires, that every valid solution of 
a microphysical problem can be referred to physical properties of
particle waves. In cases, where the established theory provides a
solution yielding, for example, correct eigenvalues but not
wave functions which can be identified as {\em physical} waves,
the result is, from this viewpoint, questionable. Essentially,
this is not limiting nor diminishing the validity of current
results, since it extends the framework of micro physics only
to regions, which so far have remained unconsidered.  

\section{Outlook}

From the five tasks defined in the introduction as our personal
criterium for a promising new approach, three have been 
accomplished by this publication. But as quantum theory, historically,
was invented due to the puzzling results in atomic physics,
the whole concept as such would be useless without a theory of
atoms compatible with the framework developed. This theory has
partly been published already (see \cite{HOF96}), partly it is
still in the making although the essential results have been 
derived. It will be the aim of the next paper published in the 
internet, to give a thorough and up to date account of a model 
of hydrogen atoms in terms of material waves.



\end{document}